\documentclass[twocolumn,superscriptaddress]{revtex4}
\usepackage{hyperref}
\usepackage{color}
\usepackage{graphics,graphicx,color}
\usepackage{bm}% bold math
\usepackage{amsmath}
\usepackage{amssymb}
\usepackage{amsfonts}
\usepackage{ifthen}

\def\(({\left(}
\def\)){\right)}                       
\def\[[{\left[}
\def\]]{\right]}

\DeclareMathOperator*{\argmin}{\arg\!\min}

\newcommand{\av}[1]{\left\langle{#1}\right\rangle}

\newcommand{\ud}{\mathrm{d}}

\definecolor{orange}{RGB}{255,127,0}
\definecolor{darkgreen}{RGB}{0,150,100}

\newcommand{\format}{pnasfigintext}  % in text
\newcommand{\onlycaption}{false}
\newcommand{\ispnas}{false}

\newcommand{\panelone}{
\begin{figure}
\centering
\ifthenelse{\equal{\onlycaption}{true}}{}{
\noindent\includegraphics[width=.9\linewidth]{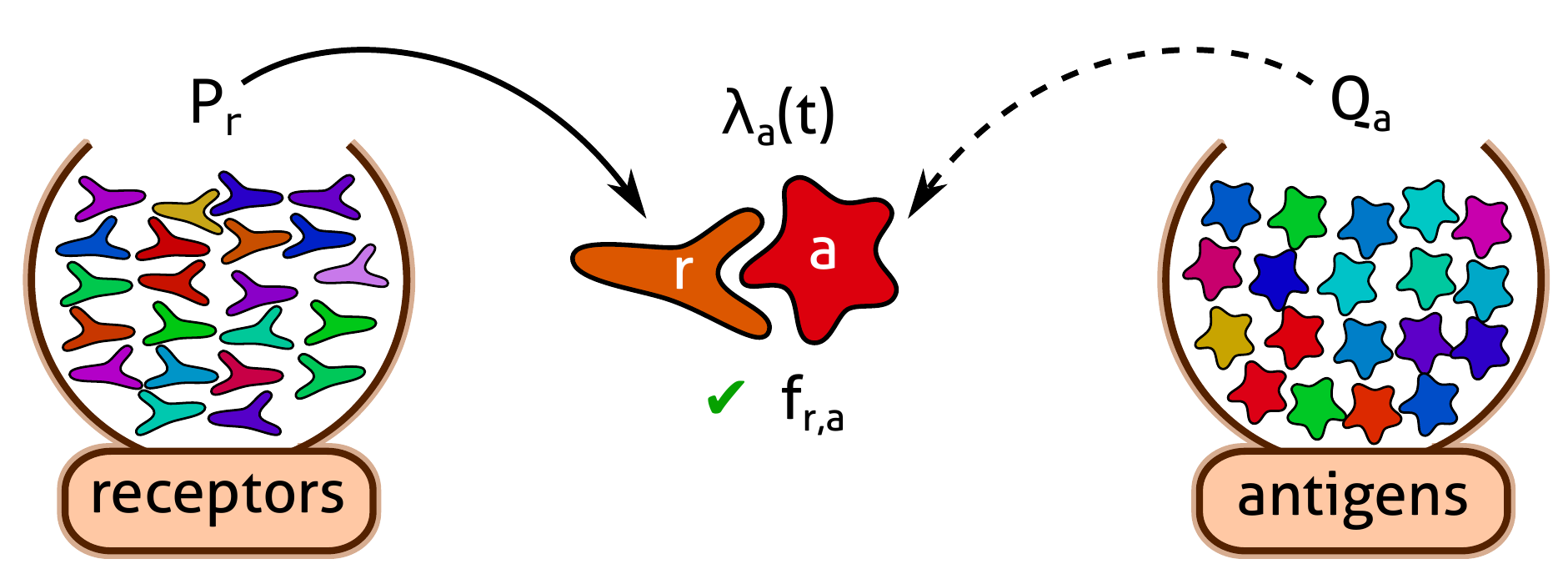}
}
\caption{Schematic of a statistical model of antigen recognition by the adaptive immune system. 
After infection, antigen $a$ encounters immune receptor $r$ at random with a rate $\lambda_a(t)$.
An encounter leads to a successful recognition with a probability
$f_{r,a}$ that reflects the matching between a given antigen--receptor
pair.
\label{fig:setup}
}
\end{figure}
}

\newcommand{\paneltwo}{
\begin{figure}
    \centering
\ifthenelse{\equal{\onlycaption}{true}}{}{
\noindent\includegraphics[width=.8\linewidth]{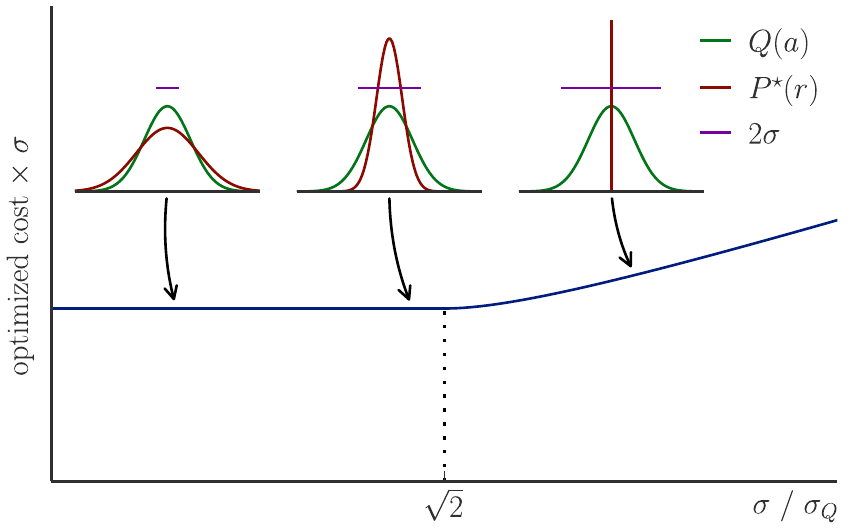}
}
    \caption{
       The optimal cost and receptor distributions for protecting against
a one-dimensional Gaussian antigenic landscape $Q(a)$ of variance $\sigma_Q^2$,
as a function of the cross-reactivity width $\sigma$.
As $\sigma$ increases, the optimal distribution $P^*(r)$ becomes narrower and narrower (left and middle insets), until it concentrates entirely onto a single point, for $\sigma\geq \sqrt{2}\sigma_Q$ (right inset).
        The minimal cost (multiplied by $\sigma$ for a comparison at
        constant recognition capability) 
is constant below the transition point, but increases with $\sigma$ past it.
The cross-reactivity function, which quantifies the affinity between receptor $r$ and antigen $a$ as a function of their distance in shape space, has a Gaussian form:
       $f(r-a) = \exp[-(r-a)^2 / 2\sigma^2]$,
       and the cost function is linear in the effective recognition
       time, $F(m) = m$.
   \label{fig:crossreactivityanalytical}
}
\end{figure}
}

\newcommand{\panelthree}{
\begin{figure}
    \centering
\ifthenelse{\equal{\onlycaption}{true}}{}{
\noindent\includegraphics[width=\linewidth]{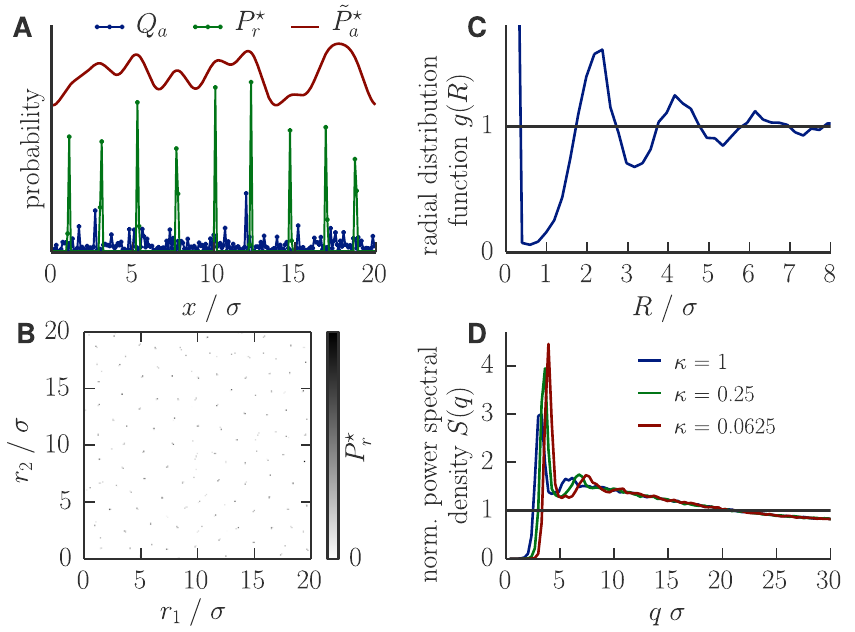}
}
    \caption{Cross-reactivity plays an important role in shaping the optimal repertoire, often leading to highly peaked repertoires.        
      (A)-(B): The optimal receptor distribution $P^*_r$ 
      for (A) one- and (B) two-dimensional random environments.
Despite being peaked, the optimal distribution of receptors covers the antigenic space fairly uniformly, as shown by its coverage by the receptors, $\tilde P^*_a=\sum_r f_{r,a}P^*_r$, shown in the one-dimensional case (A).
The cross-reactivity
and cost functions are the same as in Fig.~\ref{fig:crossreactivityanalytical}.
The antigenic landscape $Q_a$ is generated randomly from a log-normal
distribution with coefficient of variation $\kappa=1$.
(C)-(D): Structural analysis of the tiling pattern formed by the peaks of the optimal receptor distribution $P^*_r$, in two dimensions.
        (C) The radial distribution function of $P^*_r$ shows an exclusion zone around each peak, followed by oscillations characteristic of a local tiling pattern.
        (D) Normalized power spectral density $S(q)$ of $P^*_r$ for different values of the parameter $\kappa$ quantifying the heterogeneity of the antigenic landscape. The high suppression of fluctuations at large scales (small $q$) indicates that the pattern has very little fluctuations in the number of receptors used to cover large surface areas.
   \label{fig:crossreactivity}
}
\end{figure}
}

\newcommand{\panelfour}{
\begin{figure}
    \centering
    \ifthenelse{\equal{\onlycaption}{true}}{}{
\noindent\includegraphics[width=\linewidth]{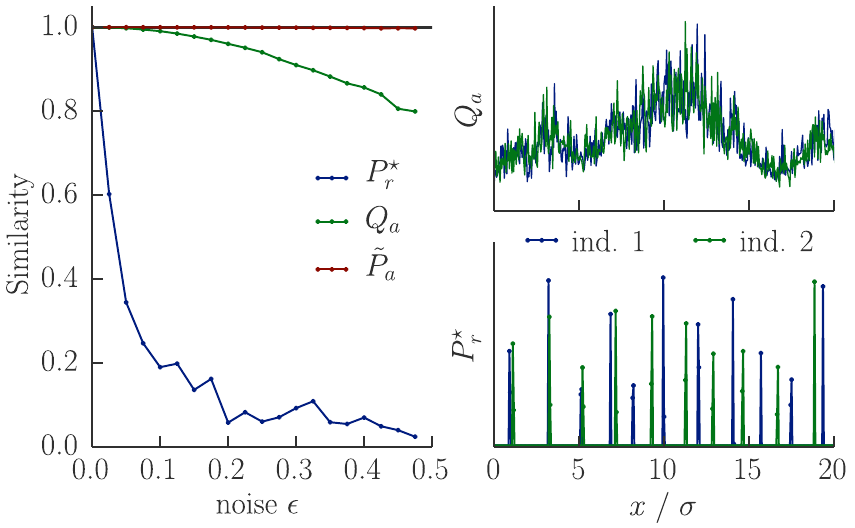}
}
    \caption{
      Two individuals in the same environment $Q_a$ that see
      it with slightly different noises have similar coverages of the
      antigenic space, but achieve it with different receptors. This
      results in largely
      non-overlapping repertoires. 
      Shown are the overlaps (normalized to be between 0 and 1) between the
      experienced pathogen distributions $Q_a$, the resulting optimal
      receptor distributions $P^*_r$, and the corresponding coverages
      $\tilde P_a$, as a function of the noise $\epsilon$ with which individuals
      perceive the environment. The right plots show an example of
      antigenic environments and optimal
      receptor distributions for $\epsilon=0.2$.
We calculated the optimal receptor distributions for two individuals 1
and 2 experiencing respective environments $Qe^{z_{1}}$ and $Qe^{z_2}$, where $Q$ is a random environment with
 fluctuations on scales larger than the cross-reactivity $\sigma$ (power spectrum $\propto 1 / (1 + (10q\sigma)^2)$) normalized so that its coefficient of variation is $0.5$,
     and $z_{1}$, $z_2$ are Gaussian noises of mean zero and variance
      $\epsilon^2$. The choice of cost and cross-reactivity functions
      are the same as in Fig.~\ref{fig:crossreactivityanalytical}.    \label{fig:individuals}}
\end{figure}
}

\newcommand{\panelfive}{
\begin{figure}
    \centering
\ifthenelse{\equal{\onlycaption}{true}}{}{
\noindent\includegraphics[width=\linewidth]{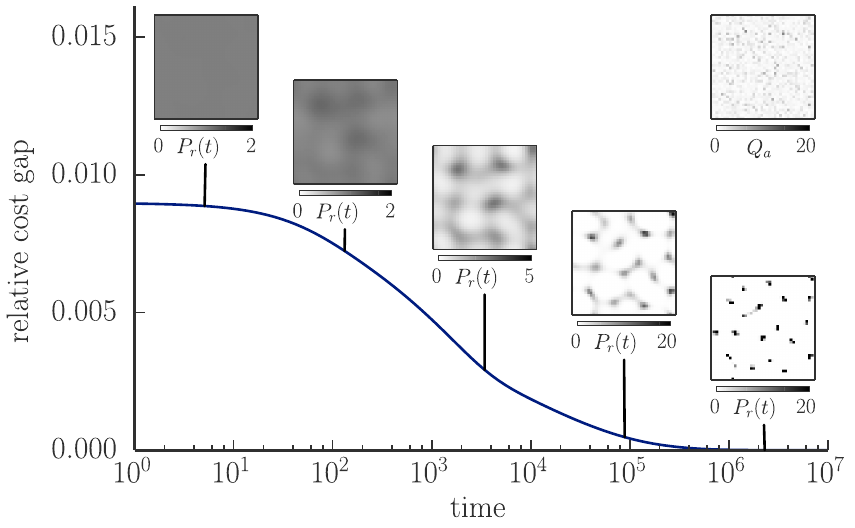}
}
    \caption{The immune repertoire can self-organize to a state that minimizes cost and provides  protection against infections via competitive evolution of receptor populations stimulated by antigens.
 Numerical solution of the population dynamics  (Eq.~\ref{eq:popdyn}) shows how competition causes a uniform initial receptor distribution to fragment into a highly peaked pattern (insets representing $P_r(t)=N_r(t)/\sum_{r'}N_{r'}(t)$).
The top-right inset represents the antigenic environment $Q_a$ driving
the dynamics (generated as in Fig.~\ref{fig:crossreactivity}B).
Departure from optimality,  as measured by the relative cost gap $[\av{F} (P_r(t)) - \av{F} (P^*_r)]/\av{F}(P_r^*)$, decreases with time and eventually reaches zero.
We use the availability function $A(\tilde N) = 1/(1+\tilde N/N_0)^2$ with $N_0=10^6$, a death rate
$d = 0.01$ and a cost function $F(m) = 1-e^{-\beta m}$ with
$\beta=0.04$. The space size is $10\sigma$.
The initial condition is uniform with $\sum_r N_r(0) = 2.5\cdot 10^7$.
    \label{fig:soo}
    }
\end{figure}
}

\newcommand{\tablecaption}{
\caption{\small\rm\sf
    The cost function $F(m)$ measures the harm caused to an organism by the time  that immune receptors have had $m$ encounters with a pathogen.   The optimal receptor distribution $P^*$  is determined by minimizing this cost, given a  pathogen distribution $Q$, and a cross-reactivity function $f_{r,a}$ specifying the probability that receptor $r$ binds to antigen $a$.      The second column gives the form of $P^*$  over scales larger than the cross-reactivity.  
  The optimal $P^*$ can be reached as a steady-state resulting from
   competitive binding between receptors and antigens (see last
   section of Results) quantified by an ``availability function'' $A$.   
$\tilde{N}_a  =   \sum_r N_r \, f_{r,a}$ represents the coverage of
antigen $a$ by the repertoire,    $N_{{\rm st}}=\sum_rN_r$ is the
total steady state population and  $C, C', \beta, {\rm and}\,  m_0$
are positive constants.
    \label{tab:costinfluence}
}
}

\newcommand{\tableone}{
\begin{table}[ht]
\ifthenelse{\equal{\ispnas}{true}}{\tablecaption}{}
   \begin{center}
    \begin{tabular}{c|c|c}%\toprule
        $F(m)$ & $P_r^*$ & $A(\tilde{N}_a)$ \\
    \hline
    $m^\alpha$ & $ C \, Q_r^{\frac{1}{1+\alpha}}$ & $C' (N_{{\rm st}} /\tilde{N}_a)^{1+\alpha}$\\
    $\ln m$ & $ C \, Q_r$ & $C' (N_{{\rm st}}/\tilde{N}_a)$\\
    $1 - \exp \((- \beta m\))$ & $\max\{{C \sqrt{Q_r} - \beta, 0 \}}$ & $C'/(\beta+ \tilde{N}_a/N_{{\rm st}})^2$\\
    $\Theta(m-m_0)$ & $\max\{{\ln\((Q_r\)) / m_0 - C, 0 \}}$ & $C'\exp(-m_0 \tilde{N}_a/N_{{\rm st}})$\\
   \end{tabular}
\end{center}
\ifthenelse{\equal{\ispnas}{true}}{}{\tablecaption}
\end{table}
}

\begin{document}

\begin{abstract}
    The repertoire of lymphocyte receptors in the adaptive immune system protects organisms from diverse pathogens.   A well-adapted  repertoire should be tuned to the pathogenic environment to reduce the cost of infections.  We develop a general framework for predicting the optimal repertoire  that minimizes the cost of infections contracted from a given distribution of pathogens.    The theory predicts that the immune system will have more receptors for rare antigens than expected from the frequency of encounters;  
individuals exposed to the same infections will have sparse repertoires that are largely different, but nevertheless exploit cross-reactivity to provide the same coverage of antigens;
and the optimal repertoires can be reached via the dynamics of competitive binding of antigens by receptors, and selective amplification of stimulated receptors.   Our results follow from a tension between the statistics of pathogen detection, which favor a broader receptor distribution, and the effects of cross-reactivity, which tend to concentrate the optimal repertoire onto a few highly abundant clones.  Our predictions can be tested in high throughput surveys of receptor and pathogen diversity.

\end{abstract}

%\title{Design principles of an optimal adaptive immune system}
\title{How a well-adapted immune system is organized}
\author{Andreas Mayer}
\affiliation{Laboratoire de physique th\'eorique,
    CNRS, UPMC and \'Ecole normale sup\'erieure, 24, rue Lhomond,
    75005 Paris, France}
\author{Vijay Balasubramanian},
\affiliation{Department of
    Physics and Astronomy, University of Pennsylvania, Philadelphia,
    PA 19104;\\
    and Initiative for the Theoretical Sciences, The
    Graduate Center, The City University of New York, NY 10016, USA}
\author{Thierry Mora}
\affiliation{Laboratoire de physique statistique,
    CNRS, UPMC and \'Ecole normale sup\'erieure, 24, rue Lhomond,
    75005 Paris, France}
\author{Aleksandra M. Walczak}
\affiliation{Laboratoire de physique th\'eorique,
    CNRS, UPMC and \'Ecole normale sup\'erieure, 24, rue Lhomond,
    75005 Paris, France}
\date{\today}

\maketitle

%immune receptor competition

 The adaptive immune system protects organisms from a great variety of pathogens by maintaining a  population of specialized cells, each specific to particular challenges.   Together these cells cover the array of potential threats.    To recognize pathogens, the  immune system relies on  receptor proteins expressed on the surface of its main constituents, the B and T lymphocytes.  These receptors interact with antigens (small molecular elements 
 making up pathogens), recognize them through specific binding, and initiate the immune response.
 Each lymphocyte expresses a unique  receptor formed from random combinations encoded in the genome.  The receptors later undergo selection through the death and division of the lymphocytes that express them, as well as mutations in the case of B lymphocytes. The diversity of the receptor repertoire  determines the range of threats that  the adaptive immune system can target.  
 
The detailed {\it composition} of the immune receptor repertoire, and not just its breadth, is important for conferring effective protection against infections.    Broadly speaking, a diverse population of receptors will confer wider immunity, and a larger clonal population of a particular receptor will confer more effective immunity against the pathogens to which it is specific.  However, there is a tradeoff between  diversity and clone sizes because the number of receptors is limited.
  By selectively proliferating some receptors at the expense of others, the immune system  retains a memory of past infections \cite{Burnet}, facilitating  subsequent immune responses.
Furthermore, while infections increase the populations of  receptors with the greatest specificity, they can also lead to a reorganization of the immune repertoire as a whole \cite{chain-2014}.

How should the repertoire be organized to minimize the cost of infections?   We develop a  framework for answering this question by abstracting key general features of the immune system: the receptor repertoire is bounded in size, receptors are ``cross-reactive'' (each antigen binds many receptors; each receptor binds many antigens), and the cost of an infection  increases with time.   Given these general assumptions, we consider a simplified  landscape of pathogens, where infections are drawn from a fixed distribution.    By simplifying the setting in this way, and independently of the detailed dynamics of immune responses,  we arrive at  broad insights about the composition of immune repertoires that are optimal for their pathogenic environments.

The theory predicts, counter-intuitively, that the number of receptors specific to rare pathogens will be amplified relative to the probability of encounter, at the expense of receptors for common infections.  We also find that two organisms responding to a pathogen distribution will display unique populations of immune receptors, even though their coverage of pathogens will be similar.   How can the immune system achieve these sorts of optima? Surprisingly, we find that simple competition between receptor clones can drive the  population to the optimal composition for minimizing the cost of infections.

 New high throughput methods are making it possible to survey B-cell and T-cell receptor diversity  in fish \cite{quake-2009,Mora:2010p5398}, in mice \cite{friedman-2012, chain-2014} and humans \cite{greenberg-2012, robins-2011, carlson-2009, mamedov-2014}.    As methods are developed to better characterize pathogenic landscapes and receptor cross-reactivity,  predictions for the composition of optimal repertoires derived from our framework can be directly compared with experiments.   To arrive at our results we ask how the immune system should be organized to  perform its function well, rather than starting with the detailed dynamics of its components.  We are proposing that the universal features of the adaptive immune system follow simply from general statistical considerations, while the detailed dynamical implementation arises from the historical contingencies of evolution.

\section{Definition of the problem}
 
%%%%%%%%%%%%%%%%%%%%%%%%%%%%%%%%%%%%%%%%%%%%%%
% July 21 version
%%%%%%%%%%%%%%%%%%%%%%%%%%%%%%%%%%%%%%%%%%%%%%% 

To find the optimal repertoire distribution we must consider the  nature of antigen-receptor interactions,
and a  penalty that the immune system pays for not recognizing antigens.
This penalty must reflect the facts that recognition should happen within reasonable time, before the pathogen colony can significantly increase its size; the interactions between the immune receptors and antigen are probabilistic; and not all antigens are equally frequent. We assume that, although the immune system cannot predict precisely which antigens it will encounter and when, it incorporates an estimate of the {\em probabilities} of their occurrences.
We also take these probabilities to be constant in time. This is an idealization grounded in a separation of timescales, which assumes the distribution of antigens remains constant on timescales on which the immune system adapts.    
%Of course, the real immune systems only has limited information, and the pathogenic environment changes constantly because of seasons, co-evolution with the host, etc. 
%For now, we will also neglect the physical and molecular basis of the elements encoding information about the antigenic environment ---\,memory lymphocytes, antibodies\,--- and the  processes by which this information is acquired and processed.}

We call $Q_a$ the probability that the next infection will be caused by antigen $a$ (Fig.~\ref{fig:setup}) and model the immune repertoire by a  distribution of receptors $P_{r}$, from which lymphocytes with the corresponding receptor are drawn at random. During its time in the periphery, an antigen $a$ will encounter and possibly interact with receptors at a rate  $\lambda_a(t)$ which  increases with time as the pathogen population grows.   Each encounter will occur with a different receptor $r$ drawn from $P_{r}$.  The mean number of encounters between  antigens  and receptors after a time $t$, which we will call {\em effective time}, is defined as
%\begin{equation}\label{noofencounter}
$m_a(t)= \int_0^t \ud \tau \lambda_a(\tau)$,
%\end{equation}
where $t=0$ is set by the introduction of the antigen.    We shall see that the cost of an infection is most easily expressed in terms of the expected number of encounters before recognition, and hence in terms of $m_a(t)$.

An antigen $a$ and a receptor $r$ interact with a certain strength set by the binding affinity between the two molecules.
This is described by the probability $f_{r,a}$ that an antigen $a$ colliding with the receptor $r$ results in a recognition event, leading to the activation of the lymphocyte expressing that receptor. $f_{r,a}$ will be called the cross-reactivity function.
Each encounter with a random antigen has a probability
%\begin{equation} \label{eq:coverage}
   $\tilde P_a = \sum_r f_{r,a} P_r$
%\end{equation}
to lead to recognition and trigger an immune response.
Since recognition is a stochastic event,
the time $t$ to the first recognition event, or response time, is random and distributed according to the probability distribution function
%\begin{equation}
  $H_a(t) = \lambda_a(t) \tilde P_a e^{-m_a(t) \tilde P_a} $  (see \ifthenelse{\equal{\ispnas}{true}}{SI Sec.~A}{App.~\ref{app:firstrec}} for a derivation).
%\end{equation}

\ifthenelse{\equal{\format}{pnasfigend}}{}{\panelone}

The longer the system fails to detect the antigen, the more likely the infection is to become harmful.
We assume that the integrated harm caused by an antigen since the beginning of an infection is an increasing function $F_a(t)$ of the time of first recognition.
The mean harm inflicted to the organism by the attack of an antigen $a$ is then given by this quantity averaged over the distribution of possible response times:
%\begin{equation}
$\bar F_a = \int_0^{+\infty} dt \, F_a(t) H_a(t) 
                                  = \tilde P_a \int_0^{+\infty} dm\, F_a[t_a(m)] \, e^{-m\tilde P_a}$,
% \label{eq:avFp}
%\end{equation}
where $t_a(m)$, the inverse function of $m_a(t)$, is the amount of time it takes for $m$ encounters to occur between the immune receptors and pathogen $a$. The result depends on the cost expressed as a function of the effective time $m$, $F_a[t_a(m)]$, which we denote $F_a(m)$ to simplify notations.
% to avoid proliferating notations.
%on the left side of the equation is understood as the composition of $F_a$ and $t_a$ on the right side.  
%$F_a(m)$ expresses the harm caused by a growing  population of pathogen $a$ by the time there are $m$ encounters.  
%Of course, the faster the pathogen population grows, the quicker it causes the same amount of harm.

We will consider several specific choices of the effective cost function in  Results.
Since not all antigens are equally likely, the overall expected cost is this harm averaged over the antigen distribution:
\begin{equation} \label{eq:cost}
   \mathrm{Cost}(\{P_r\})=\langle F\rangle=\sum_a Q_a \bar F_a.
\end{equation}
The need to defend against many antigens at the same time with a limited number of receptors introduces a trade-off.
If more receptors recognize an antigen, there are less  to protect against other threats.

Our aim here is to propose a general framework for thinking about the repertoire.  Thus, we do not explicitly model  intracellular communication, cell differentiation, activation of co-factors, coordination of different cell types,  avoidance of self-antigens through thymic selection, and the full complexity of the recognition process.  The idea is that $F_a(m)$ implicitly summarizes all of these factors in terms of an effective cost.
Of course, more detailed modeling of the cost will be possible as we refine our knowledge of the recognition process.

In general the cost function $F_a(m)$ depends on the antigen $a$, reflecting the various virulences of different pathogens. To simplify, we can assume that the cost function takes the factorized form: $F_a(m)=\mu_a F(m)$, where $\mu_a$ is the pathogen-dependent virulence factor, and $F(m)$ describes how all threats develop with time. The cost will then take the form: $\sum_a \mu_a Q_a \tilde P_a \int_0^{\infty} dm F(m)e^{-m\tilde P_a}$. In this expression, the virulence factor $\mu_a$ of a pathogen plays the same role as its likelihood $Q_a$. Some pathogens are rare but very virulent (like anthrax), while others may be common but not very virulent (like the common cold), and an ideal immune system should be able to cope with both. In our model the overall ``dangerousness'' of a pathogen is expressed as the product of the two, $\mu_aQ_a$. Therefore, for all practical purposes $\mu_a$ can be absorbed into the definition of $Q_a$, and will be omitted in the rest of the paper.

Given such a model of the recognition process, there exists an optimal adaptive immune system, { characterized by the choice of the receptor distribution $P_r$,} that minimizes the expected cost in a given antigenic environment $Q_a$.
The optimal repertoire is found by minimizing the expected cost in Eq.~\ref{eq:cost} with respect to $P_r$, subject to constraints of non-negativity ($P_r \geq 0$) and normalization ($\sum_r P_r =1$). Simple local extremality conditions are sufficient for optimality because our problem can be shown to be convex (see \ifthenelse{\equal{\ispnas}{true}}{SI Sec.~B}{App.~\ref{app:convexity}}).   The condition $\sum_r P_r =1$ is a normalized version of the constraint that the total number of receptors is limited.

\section{Results}
\subsection{The optimal repertoire is more uniform than the pathogen distribution}\label{sec:res1}
We can now ask how best to distribute the receptors to minimize the cost  (Eq.~\ref{eq:cost}) for a given antigenic environment.
To begin, we neglect cross-reactivity (later we will see that this is equivalent to looking at the structure of the repertoire at scales larger than the cross-reactivity).
In this case antigens and receptors can be associated one by one by a  cross-reactivity function $f_{r,a} = 1$ if $r=a$ and $0$ otherwise.  
%Because of this simplification, the only remaining coupling between the types of receptors is through the global normalization constraint $\sum_r P_r = 1$. 
 In this case we can analytically determine the optimal distribution (\ifthenelse{\equal{\ispnas}{true}}{SI Sec.~D2}{App.~\ref{app:optspecific}}):
\begin{equation}\label{eq:bestPr}
P^*_r=\max\left[\bar F'^{(-1)}\left(-{\frac{\lambda}{Q_r} } \right),0\right],
\end{equation}
where $\bar F'^{(-1)}$ denotes the inverse function of the derivative of $\bar F_a=\bar F(\tilde P_a)$ expressed as a function of $\tilde P_a$, and $\lambda$  is a positive constant fixed by the normalization $\sum_r P^*_r=1$.  Table~\ref{tab:costinfluence} presents results for several representative cost functions.

A simple scenario occurs when the pathogen population grows exponentially in time, as do the cost and the encounter rate---reflecting the proliferative nature of pathogens. In this case the cost grows linearly in the number of encounters, {\em i.e.} $F(m) = m$ (see \ifthenelse{\equal{\ispnas}{true}}{SI Sec.~C}{App.~\ref{app:costexample}}).  Then we find that the optimal fraction of the repertoire taken up by a given receptor is proportional to the square root of the frequency of the corresponding antigen $P_r^* \propto \sqrt{Q_r}$. Intuitively, we expect that the optimal repertoire should focus its resources on receptors recognizing the most common antigens.
%For these common infections, the immune system will respond more quickly compared to one with a uniform distribution of receptors.
However this enhanced protection against frequent antigens comes at the cost of a slower response against the uncommon antigens, and this bias towards common antigens must remain limited. The square root dependence reflects a particular trade-off between these two opposing constraints, by directing more resources towards common antigens while uniformizing the distribution compared to a linear dependence.  Intriguingly, the same square root dependence has been found as an optimal solution for the size of tRNA pools as a function of codon usage \cite{xia-1998}, and
 in a model for the screening of suspicious individuals \cite{press-2009}.

The extent to which more resources are directed towards common antigens depends on the relative gains and losses of earlier and later recognition events, which are captured in our model by the effective cost function $F(m)$.
In general,
steeper cost functions {imply} more flattened  distributions of receptors.
The cost function $F(m)=m^\alpha$, and its associated optimal distribution $P^*_r\propto Q_r^{1/(1+\alpha)}$, help illustrate this point. Such cost functions can arise when both $m(t)$ and $F(t)$ increase exponentially as a function of time, but with different exponents
%They may also originate from the cooperative nature of pathogenic action
(see \ifthenelse{\equal{\ispnas}{true}}{SI Sec.~C}{App.~\ref{app:costexample}}).
When $\alpha$ is large, the cost of non recognition increases very quickly with time, calling for an urgent response. Consequently the optimal immune system tends to cover the space uniformly to get all potential threats, even the unlikely ones, under control. Conversely, when $\alpha$ is low, the harm caused by pathogens does not explode with time, meaning that the system can afford to recognize the rarer pathogens late, and focus its resources on the common ones.

In some situations, there may be little or even no difference between a late response, or no response at all, because the total harm caused by an infection stabilizes.  For example, consider the cost  $F(m) = 1 - e^{-\beta m}$ which saturates at large effective times.    In this case,
the optimal solution (Table~\ref{tab:costinfluence}) relates receptor and antigen through a square root  as for linear cost, but with a cut-off at low probabilities.  This cut-off occurs because there is little benefit to having receptors recognizing rare antigens, whose recognition is likely to happen late,
when differences in recognition times do not matter anymore. 
 
\ifthenelse{\equal{\ispnas}{true}}{}{\tableone}

Real harm 
may occur only when the effective time $m$ crosses a threshold. This situation can be modeled by taking  $F(m) = \Theta(m-m_0)=0$ for $m<m_0$, and $1$ otherwise. In this case the receptor distribution should be organized to maximize the chance of detection before $m_0$.
The optimal repertoire for this cost (Table~\ref{tab:costinfluence}) has no receptors for the least frequent pathogens (cutoff at low probabilities) and a drastically flattened receptor distribution (logarithm of the pathogen distribution). 
 
Is there a cost function for which the receptor distribution is not flattened relative to the pathogen distribution?   This occurs in a special case where cost increases very slowly (logarithmically) with effective time.    However, in general, cost is minimized by a receptor distribution that is flattened relative to the pathogen distribution.

\subsection{Cross-reactivity dramatically reduces diversity in the optimal repertoire}
\label{sec:res2}

By allowing receptors to bind to a variety of  antigens, cross-reactivity should permit the immune system to reduce the number of receptor types required to cover the whole range of possible threats.   We will show that given sufficient cross-reactivity, the optimal immune repertoire concentrates all its resources on a few receptors, which together tile antigenic space.

Following Perelson and Oster \cite{oster-1979},  we think of receptors and antigens as points in a common high dimensional {\it shape space}, whose coordinates are associated to unspecified physicochemical properties. For simplicity,  assume that cross-reactivity only depends on the relative position of receptor and antigen in shape space $f_{r,a} = f(r-a)$,  where $f$ is a decreasing function of the distance between $a$ and $r$.
Short distances in shape space correspond to a good fit between the two molecules, leading to strong recognition, while large distances translate into weak interactions and poor recognition.

In order to build  intuition, we first consider an analytically solvable example (Fig.~\ref{fig:crossreactivityanalytical}).  
We describe the space of receptors and antigens by a single continuous number, and
assume a Gaussian antigen distribution with variance $\sigma_Q^2$, and Gaussian cross-reactivity of width $\sigma$,
which sets the typical distance within which a receptor and  antigen interact.
We derive the optimal receptor distributions  analytically for costs of the form $F(m) = m^\alpha$
(\ifthenelse{\equal{\ispnas}{true}}{see SI, Sec. D3b}{App.~\ref{app:gaussian}}).
For narrow cross-reactivities ($\sigma < \sigma_c =  \sigma_Q\sqrt{1+\alpha}$), the optimal receptor distribution is  Gaussian with variance $(1+\alpha)\sigma_Q^2 - \sigma^2$
and the optimal cost is independent of  $\sigma$.   For wide cross-reactivities ($\sigma > \sigma_c$),  the receptors are optimally of a single type with reactivity centered on the pathogen distribution,  while the optimal normalized cost increases with $\sigma$ since the receptor is unnecessarily broadly reactive.
  These results arise from a tension between two opposing tendencies.  As in the non cross-reactive case,
the need to cover rare pathogens
broadens
the optimal receptor distribution relative to the pathogen distribution. 
  But cross-reactivity has the opposite effect, favoring more concentrated distributions.

\ifthenelse{\equal{\format}{pnasfigend}}{}{\paneltwo}

\begin{figure}
    \centering
    \includegraphics{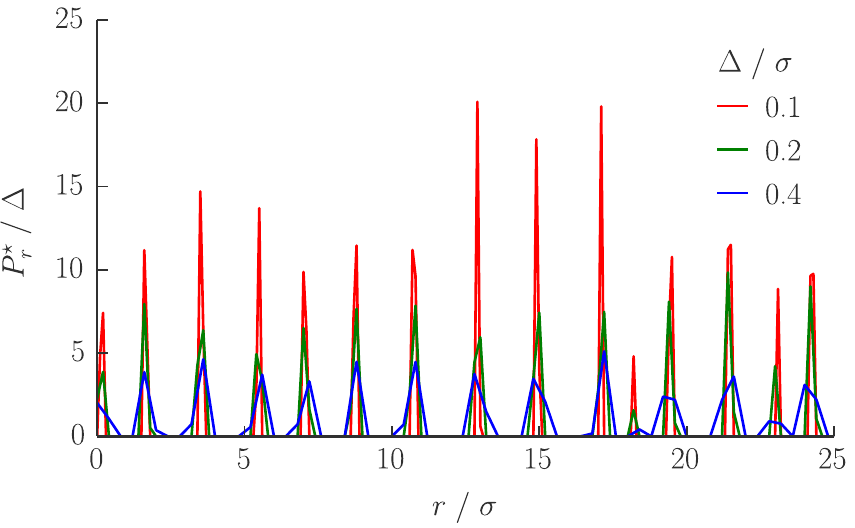}
    \caption{
Solving the optimization problem with a finer and finer discretization step
suggests that the peaks found in the optimal receptor distributions
converge to true Dirac delta functions. Starting from
a problem with a discretization step of $\Delta=0.1\sigma$, we
construct coarse-grained versions of it
by downsampling the antigen distribution two and four fold, yielding
$\Delta =0.2$ and $0.4$ respectively. The resulting
coarse-grained optimization problems are then solved, and the optimal
distributions $P_r^*/\Delta$ represented (after appropriate normalization by the
step size). The random antigen distribution is log-normal with
coefficient of variation $\kappa=0.25$.
%        Exemplary cutouts of optimal repertoires for a series of downsampled optimization problems shows convergence of the peaks in the receptor distribution against Dirac delta functions.
%        We consider a one dimensional receptor -- antigen space with periodic boundary conditions and express all lengths relative to the box length $B$.
%        Cross-reactivity is described by $f(r-p) = \exp\((-|r-p|^2 / 2\sigma^2\))$ with $\sigma = 0.01 B$.
%        The original problem consists in finding the optimal receptor distribution for a random antigen landscapes $Q \propto \ln \mathcal{N}(0.25)$ on a grid with spacing $\Delta = 0.001B$.
%        Related problems at lower discretization $\Delta$ are then created by down-sampling the antigen distribution with a Fourier space technique keeping only the lowest frequencies.
    \label{fig:dirac}
    }
\end{figure}

Does cross-reactivity generically drive the optimal receptor distribution to cluster into peaks?  We investigated
 this question numerically.
For concreteness, we consider a linear cost $F(m) = m$, and random pathogen environments in one or two dimensions constructed by drawing each $Q_a$ from a log-normal distribution characterized by a coefficient of variation $\kappa$.
For numerical purposes, the shape space is taken to be bounded and discretized, and we use  accelerated gradient projection optimization (\ifthenelse{\equal{\ispnas}{true}}{SI Sec. E}{App.~\ref{app:numerics}}). We find that the optimal repertoire $P^*$ is strongly peaked on a discrete forest of receptors  (Fig.~\ref{fig:crossreactivity}A,B).  The width of these peaks decreases as numerical precision is increased, suggesting that the true optimum consists of a weighted sum of Dirac delta functions, {\em i.e.}  distinct, discretely spaced receptors in different amounts (see \ifthenelse{\equal{\ispnas}{true}}{SI Fig.~S1}{Fig.~\ref{fig:dirac}}).    By inspection, the peaks are spaced  evenly, at  roughly
the cross-reactivity scale $\sigma$, suggesting that $P^*$ is smooth when viewed at scales larger than $\sigma$.  Confirming this, $\tilde{P}^*$ ({\em i.e.} 
the coverage of the antigenic space by the receptors)
smoothly tracks the variations in the antigen distribution $Q_a$ at a broad scale (Fig.~\ref{fig:crossreactivity}A).   When viewed coarsely in this way, cross-reactivity is irrelevant and $P^*$ tends to the solutions of Table~\ref{tab:costinfluence}.

\ifthenelse{\equal{\format}{pnasfigend}}{}{\panelthree}

How can we quantitatively understand such distributions which are fragmentary in detail, but show organization when viewed coarsely (Fig.~\ref{fig:crossreactivity}B)?   A useful technique, borrowed from condensed matter physics, is to measure the {\em radial distribution function}  \cite{ChaikinLubensky}:
  $ g(R) = \av{P(r) P(r')}_{|r - r'|=R}$,
where $|r-r'|$ is the distance between points $r$ and $r'$.
Fig.~\ref{fig:crossreactivity}C presents $g(R)$ for $P^*$ in two dimensions. The initial drop at small $r$ indicates that peaks in $P^*$ are rarely close -- {\em i.e.}, peaks in the optimal repertoire tend to repel each other.
This exclusion, which operates over the range of strong cross-reactivity, is a sensible way to distribute resources, as it limits redundant protection against the same pathogens.
The damped oscillation of the peaks of $g(R)$ confirm that the receptors in $P^*$ are organized into a disordered tiling pattern.   A similar radial distribution function is seen in high density random packings of hard spheres where the spheres must cover as much space as possible but exclude each other. In both cases, the tiling ensures uniform coverage of space at large scales.

\begin{figure}
    \centering
\noindent\includegraphics[width=\linewidth]{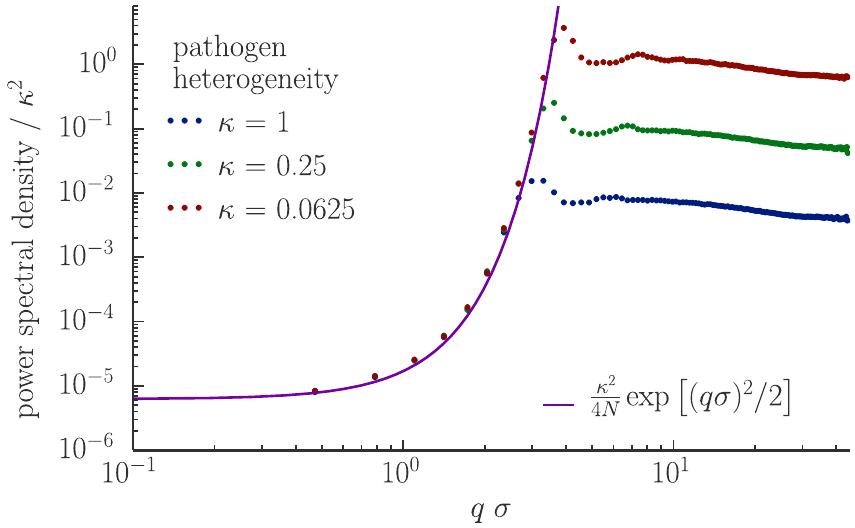}
    \caption{
Power spectral density normalized by the squared antigenic environment
heterogeneity index $\kappa$: $|\sum_r P_r e^{\mathrm{i}qr}|^2/ \kappa^2$.
The data collapse for different $\kappa$
shows that the fluctuations at large
        scale are entirely attributable to those of the antigenic
        environment, and scale with them. At these large scales, the power spectrum of the receptor distribution is approximately given by:
$\exp[(q\sigma)^2/2]/4$. The exponential term stems from the inverse
of the Fourier transform of $f$ (see
Eq.~\ref{eq:optfourier}). Parameters are the same as in Fig.~3.
    \label{fig:kappascaling}
    }
\end{figure}

To quantify the  regularity of  the tiling, we calculate the {\em normalized power spectral density} of the 2D pattern:
$S(q)={\sum_{r,r'} P_rP_{r'} e^{{\rm i} q(r-r')}}/{\sum_r P_r^2}$,
where $q$  is a wave vector. Large (small)  $|q|$ correspond to short (long) distances in antigen shape space.
When $P_r$ is made of Dirac delta peaks of uniform heights, $S(q)$ coincides with the {\em structure factor} familiar in physics, and satisfies $S(q \rightarrow \infty) = 1$. Fig.~\ref{fig:crossreactivity}D shows $S(q)$ averaged over many realizations of the antigen landscape, and over all directions of $q$ so that it only depends on its modulus $|q|$.
$S(q)$ approaches 1 for large $q$, showing that the precise local positions of the peaks are random. (The small departure from $1$ is attributable to numerical discretization.)
$S(q)$ is very low for small $q$, indicating that the number of receptors contained in any given large area of the shape space is very reproducible, providing uniform coverage.  This phenomenon of small scale randomness with large-scale regularity is called  {\em hyperuniformity}  \cite{stillinger-2003}, and arises in jammed packings \cite{Donev:2005p13220,Berthier:2011p13219} as evidence of  the incompressibility of the material.
For our optimal repertoires  small scale fluctuations (large $q$) get smoothed out by cross-reactivity and can be tolerated, while at large scales the fluctuations track variations in the antigenic landscape to provide smooth coverage (see \ifthenelse{\equal{\ispnas}{true}}{SI Fig.~S2}{Fig.~\ref{fig:kappascaling}}).

\ifthenelse{\equal{\format}{pnasfigend}}{}{\panelfour}

\begin{figure*}
    \centering
    \includegraphics{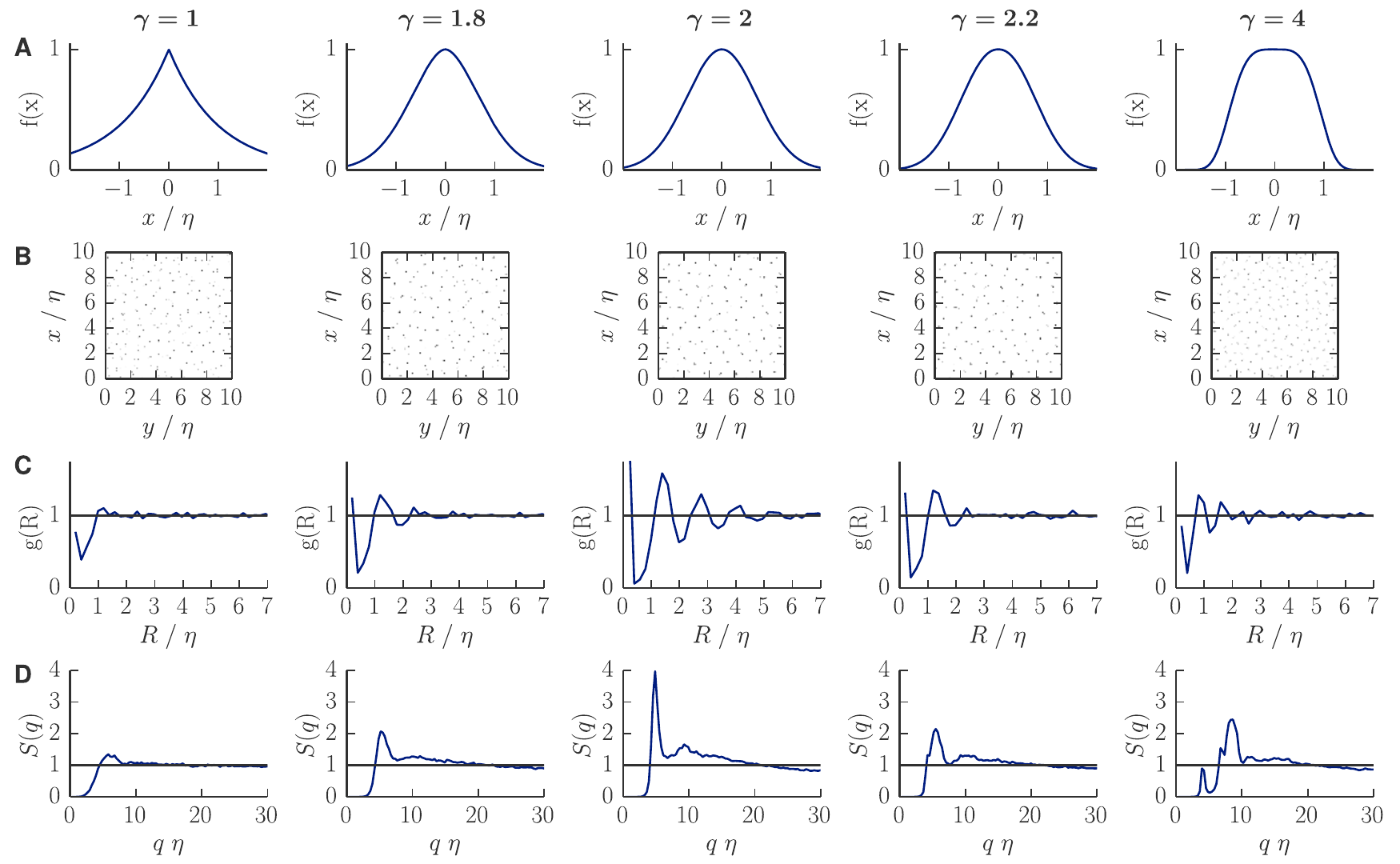}
    \caption{
        Influence of the choice of the cross-reactivity kernel $f(a-r)$ on the optimization problem.
        Regardless of the kernel choice the optimal repertoire is peaked for non-uniform antigen distributions.
        The details of distribution depend on the cross-reactivity kernel.
        (A): Kernel functions used to describe cross-reactivity.
        We use the family of kernel functions defined by $f(r-a) = \exp[-(|r-a|/\eta)^\gamma]$.
        By changing the parameter $\gamma$ we can go from an
        exponential ($\gamma = 1$) via a Gaussian $\gamma = 2$
         to a top-hat kernel ($\gamma \rightarrow \infty$).
        Up to $\gamma = 2$ all such kernels have positive Fourier transforms, whereas for $\gamma > 2$ the Fourier transforms also take negative values \citep{hernandez-garcia-2007}.
        (B): Examples of optimal receptor distributions in two
        dimensions, for antigenic environments generated as in Fig.~3B
        (with coefficient of variation $\kappa=0.25$).
% Fig.~\ref{fig:crossreactivity}B.
%        \ifthenelse{\equal{\ispnas}{true}}{Fig.~3B of the main text}{Fig.~\ref{fig:crossreactivity}B}.
%        A two dimensional receptor -- antigen space with periodic boundary conditions is used.
%        The problem is discretized on a $20\eta \times 20\eta$ grid with a discretization spacing of $0.1\eta$.
%        A random antigen landscape is generated for each run according to $Q \propto \ln \mathcal{N}(0.25)$.
        (C) Radial distribution function of the optimal distribution.
        (D) Structure factor of the optimal distribution.
        The results in both (C) and (D) are averaged over 10
        independent runs.    A linear effective cost function $F(m) =
        m$ is assumed throughout.  The random antigen distribution is log-normal with
coefficient of variation $\kappa=0.25$.
    \label{fig:kernels}
    }
\end{figure*}

\begin{figure}
    \centering
    \includegraphics{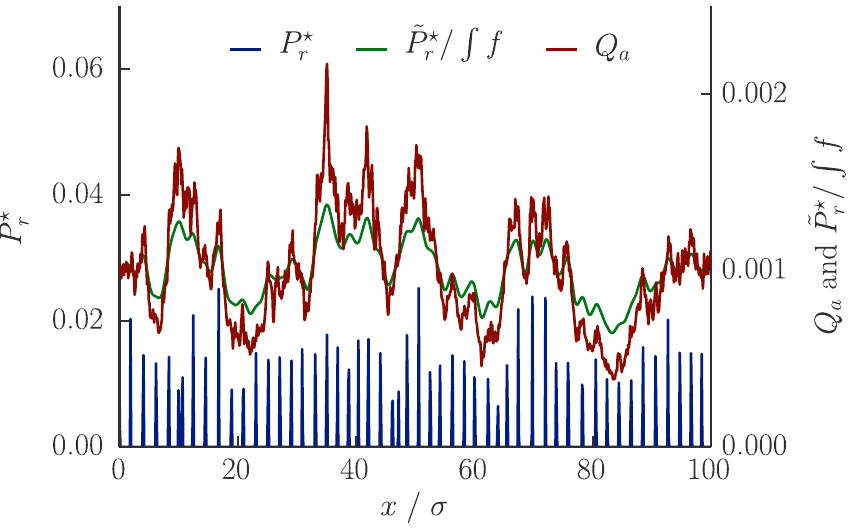}
    \caption{
        Adding correlations to the antigen distribution does not change the peakedness of optimal receptor distributions.
%        A one dimensional receptor -- antigen space with periodic boundary conditions on a box of length $B$ is assumed.
 %       The problem is discretized with a spacing of $0.0001 B$. 
 %       Cross-reactivity is described by $f(r-p) = \exp\((-|r-p|^2 / 2\sigma^2\))$ with $\sigma = 0.001 B$. 
        The result of the optimization is shown for a random antigen
        landscape with correlations.
        The antigen distribution is generated by Fourier filtering.
        First we generate an uncorrelated, normally distributed random series.
        This series is then filtered to obtain a power spectrum $\propto 1 / (1 + (10q\sigma)^2)$.
        Finally, the filtered series is exponentiated to ensure the non-negativity of the generated values.
    \label{fig:correlation}
    }
\end{figure}

To test the generality of our findings we tested other choices of cross-reactivity functions. We found that the optimal repertoire remains strongly peaked, although the position, number and strength of the peaks do change  (\ifthenelse{\equal{\ispnas}{true}}{SI Fig.~S3}{Fig.~\ref{fig:kernels}}).   Next we considered distributions of antigens with correlations across shape space (reflecting {\em e.g.} phylogenic correlations between pathogens). Again we find peaked optimal receptor distributions (Fig. \ifthenelse{\equal{\ispnas}{true}}{SI Fig.~S4}{\ref{fig:correlation}}), similar to those for  uncorrelated antigen landscapes.  For computational reasons, we restricted our analysis to two dimensional pathogen landscapes, but the analogy with random packing problems that we discussed above allows us to expect that all of these results will hold  generally in higher dimensions.

In summary, the optimal immune repertoire looks  random at scales smaller than the cross-reactivity, but has the structure of a disordered tiling at larger scales so that, after accounting for cross-reactivity, the repertoire smoothly covers the pathogen landscape.   These findings have an important consequence for different individuals exposed to the same pathogenic environment. 
  Each individual will experience a slightly different spectrum of antigens because of the statistics of encounters and other sources of variability.
These slightly different experiences of the same world lead to optimal repertoires with a striking property -- the receptor distributions are largely different, even though their coverage of the pathogen landscape is similar after including cross-reactivity (Fig.~\ref{fig:individuals}).   This finding can be compared with surveys of ``public'' repertoires of immune receptors \cite{Venturi:2008p13151,chain-2014}.

\subsection{The optimal repertoire can be reached through competition for antigens}
\label{res3}

The results presented so far have established how repertoires should be structured to provide optimal protection.
Given the complex interdependences between receptors arising from local and global trade-offs, one might think that the globally optimal solution could only be reached via some biologically implausible centralized mechanism distributing resources system-wide. In fact, we will show that the optimal repertoire can be reached through self-organization, via competitive evolution of receptor populations   under antigen stimulation.

We consider a model  that is  similar to that introduced by de Boer, Perelson and collaborators for competitive dynamics of B and T cells  \cite{perelson-1994,perelson-2001}.
Its main assumptions are that division of  receptor-expressing lymphocytes is driven by antigen stimulation, and that receptors compete for the limited supply of antigens. The number $N_r$ of receptors of a given type $r$ evolves according to:
\begin{equation} \label{eq:popdyn}
\frac{\ud N_r}{\ud t} = N_r \[[\sum_a Q_a A\((\sum_r N_r f_{r,a}\)) f_{r,a} - d \]].
\end{equation}
Receptors proliferate upon successful recognition by antigens (first term of the equation) and die with a constant rate $d$ (second term of the equation). The growth rate of a receptor type is proportional to the number of antigens that it detects. In the absence of competition, this amount is simply
 $\sum_a Q_a f_{r,a}$, but the antigen $a$ will also bind other receptors, reducing its availability for receptor $r$.
The {\em coverage} of antigen $a$ by the repertoire, $\tilde N_a = \sum_r N_r f_{r,a}$, quantifies the breadth of the receptor pool competing to bind with $a$.
The availability of antigen $a$ for binding is assumed to be a decreasing function $A(\tilde N_a)$ of its coverage. The  stimulation of $r$  by $a$ is thus modified to: $\sum_a Q_aA(\tilde N_a) f_{r,a}$ as in Eq.~\ref{eq:popdyn}.
For a given pathogenic environment, the total steady-state receptor population size $N_{{\rm }}$ will be set by the death rate $d$, which counter-balances growth at steady state.

The stable fixed points of the dynamics (\ref{eq:popdyn}) realize the optimal repertoires of the previous sections when the availability function $A$ is  matched to the cost function $F(m)$ through the relation
\begin{equation} \label{eq:condlyapunov}
    A\((\tilde N_a\))=-c'\bar F'\((\tilde N_a/N_{\rm st}\)),
\end{equation}
where $N_{\rm st}$ is the total number of receptors $\sum_r N_r$ at steady state.
Table \ref{tab:costinfluence} shows  $A(\tilde N)$ for several cost functions.
To understand  this result, first note that when binding is {\em not} cross-reactive the dynamical equations for each receptor are independent, and read: $dN_r/dt=N_r(Q_rA(N_r)-d)$. The availability function now depends only on $N_r$, meaning that receptors  only compete with their own kind ---\,they occupy their own antigenic niche. The steady state size of clone $r$ is thus set by the carrying capacity of that niche, $N_r=A^{(-1)}(d/Q_r)$, or zero if that capacity is negative.  With the availability given by Eq.~\ref{eq:condlyapunov}, this reproduces the optimal repertoire (Eq.~\ref{eq:bestPr}).
A similar argument holds when receptor binding is cross-reactive (\ifthenelse{\equal{\ispnas}{true}}{SI Sec.~F}{App.~\ref{app:popdynfixedpoint}}).
Cross-reactivity leads to competition amongst  receptor types, effectively enforcing an exclusion between similar receptors. This phenomenon, known in ecology as competitive exclusion,  is important for lymphocyte dynamics  \cite{perelson-1994}, and provides the mechanism by which our dynamical model reproduces the discrete clustering found in the optimal receptor distribution.

\begin{figure}
    \centering
    \includegraphics{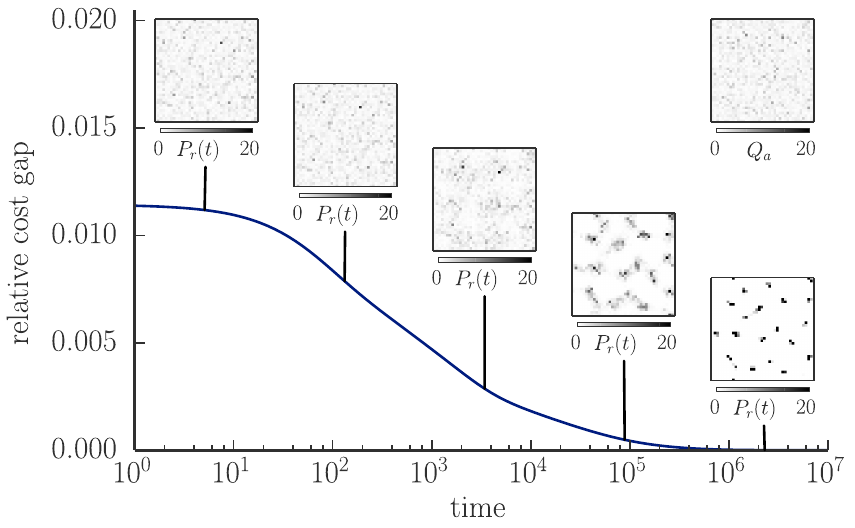}
    \caption{Numerical solution of the population dynamics described
      in the main text as a function of time. The same calculation as
      in Fig.~5, but using a random initial condition (log-normal with
      coefficient of variation 1) shows the same convergence to the optimal receptor distribution.
    \label{fig:sooapp}
}
\end{figure}

To check that the dynamics do converge to the optimum, we simulated Eq.~\ref{eq:popdyn} numerically for a random antigenic environment in two dimensions, with $A(\tilde N)=1/(1+\tilde N/N_0)^2$.
Fig.~\ref{fig:soo} shows the dynamics of the receptor distribution $P_r(t)=N_r(t)/\sum_{r'}N_{r'}(t)$, as well as its cost relative to the optimal solution, as a function of time.
Starting from a uniform initial distribution of receptors, the repertoire 
reorganizes into localized peaks that become increasingly prominent and well-separated with time, with almost  no receptors in between. 
Starting from a random initial condition leads to the same steady state (\ifthenelse{\equal{\ispnas}{true}}{SI Fig.~S5}{Fig.~\ref{fig:sooapp}}).
The cost converges towards the global minimum, indicating that the steady-state solution is indeed optimal.

\ifthenelse{\equal{\format}{pnasfigend}}{}{\panelfive}

In summary, competitive dynamics can allow the immune repertoire to self-organize into a state that confers high protection against infections.   
In the special case when the availability $A$ is scale invariant,
the expected cost is a Lyapunov function of the dynamics (\ifthenelse{\equal{\ispnas}{true}}{SI Sec.~G}{App. \ref{app:popdynlyapunov}}).  In this case, we can prove analytically  that regardless of the initial condition the cost will steadily decrease until the optimum is reached.

\section{Discussion}

We introduced a general framework for predicting the optimal  composition of the immune repertoire to minimize the cost of infections contracted from a given distribution of antigens.
This framework can be extended in several ways to be more biologically faithful, {\em e.g.} by accounting for receptor-dependent cross-reactivities, antigen-dependent infection dynamics, and evolution of the pathogenic landscape.   Our predictions can be tested in experiments  that  study how  the environment influences the composition of immune repertoires, either via  high-throughput sequencing surveys of receptor populations \cite{Vollmers:2013p13095,chain-2014}, or by sequencing receptors specific to given antigens \cite{jenkins-2007}.   The comparison between theory and experiment will provide insight into the functional constraints of antigen recognition by the immune system.

There are many situations where living systems must respond to
very diverse and often very high dimensional spaces of external influences using strictly limited resources.
To sense, internally represent, and then respond to  these influences, organisms often employ  a large diversity of components, such as cell types or genes \cite{Tkacik:2009p4346}, each sensitive to a small part of the space.  
For example, the retina supports a diverse population of ganglion cell types, each sensitive to a different visual feature, that collectively  represent the behaviorally salient aspects of visual scenes \cite{Masland, Meister}.  Likewise, the mammalian olfactory system contains some $\sim$1000 distinct receptors that each bind widely to odorants, and collectively cover olfactory space \cite{BuckAxel}.  
 In these cases,  the limited repertoire of component types provides a key constraint on information processing.  Faced with such constraints, living systems must commit resources wisely, adapting to the structure of the environment, and balancing breadth of coverage against depth of resolution, in light of priorities, costs and constraints \cite{BalasubramanianSterling}.   We have shown that these elements also shape the optimal form of the immune repertoire. 

Our finding that cross-reactivity causes the optimal repertoire to fragment is related to the concept of {\em limiting similarity}  due to competitive exclusion in ecological settings \cite{levins-1967,meszena-2006,nes-2006,hernandez-garcia-2007,dieckmann-2013}. 
In the latter context, empty regions of phenotypic space result when competition is  important on the scale at which resources vary \cite{levins-1967}, and continuous coexistence of species only occurs in exceptional cases  \cite{hernandez-garcia-2007}.  
In general, niche-space heterogeneity promotes species clustering \cite{meszena-2006,dieckmann-2013}, recalling our finding  that any heterogeneous antigen distribution leads to fragmentation of the optimal repertoire.
The conceptual connection between the immune repertoire and  ecological organization  is  even clearer in our  dynamical model where species compete for an array of resources (the antigens), and grow in relation to their success in securing  resources.

Although this study relies on a simple abstraction of the adaptive immune system, we expect that our framework and results will extend  to other distributed protection systems where diverse threats are addressed by an array of specific responses. 
For example, the immune system of bacteria, or CRISPR system \cite{Marraffini:2010p9775}, for which population dynamics models have already been proposed \cite{He:2010p9888}, could be studied within a similar framework to predict the relative abundance of CRISPR spacers and corresponding viruses in a co-evolving population of bacteria and viruses.

{\bf Acknowledgements.} The work was supported by grant ERCStG n. 306312. 
VB was supported by the Fondation Pierre-Gilles de Gennes, NSF grants PHY-1058202 and EF-0928048.  Portions of this work were done at the Aspen Center for Physics, supported by NSF grant PHY-1066293. AM was supported by a DAAD Promos stipend.

\appendix
\section{Probability distribution of the time of first recognition}
\label{app:firstrec}
In order to calculate the cost of not-recognizing an antigen $a$, we need to find the distribution of times when a successful encounter takes place.
%We will derive this probability distribution by extending the derivation of the well-known exponential distribution of waiting times characteristic of a homogeneous Poisson processes found in \cite{bialek-2012} to the thinned, inhomogeneous Poisson process considered here.
The probability of having the first recognition of antigen $a$ by
receptor $r$ in the time between $t$ and $t+\ud t$ reads: 
\begin{equation*}
    \begin{split}
        H_a(t) \ud t &= \lambda_a(t) \ud t \cdot \sum_r P_r f_{r,a}  \\
                     &\times \lim_{N\rightarrow\infty} \prod_{i=1}^N \(( 1 - \lambda_a(t_i) \frac{t}{N} \sum_r P_r f_{r,a} \)),
    \end{split}
\end{equation*}
where the first term is the probability of having an encounter between
$t$ and $t+\ud t$, the second the probability of this encounter being
successful, and the third the probability of there not being any prior
recognition events.
%{\TM expressed as a product on many infinitesimal intervals of
%  duration $t/N$.}
For the calculation of the last term we have decomposed the time leading up to $t$ into $N$ intervals of length $t/N$.
%Due to the statistical independency of the events in a Poisson process this term factorizes,
%and we use that the probability to have an event in a small window $\Delta t$ around the time $t_i = i \Delta t$ is equal to $\lambda_a(t_i) \Delta t$.
%We can rewrite this term as 
%\begin{equation*}
%    \exp\((\ln \prod_{i=1}^N \[[ 1 - \lambda_a(t_i) \Delta t \cdot \tilde{P}_a \]] \)),
%\end{equation*}
%where we have introduced the short-hand notation $\tilde{P}_a = \sum_r P_r f_{r,a}$ for the probability that a randomly chosen receptor recognizes the antigen $a$.
%Using the logarithm laws and the fact that $\ln(1-x) \approx -x$ for small $x$ we obtain
%\begin{equation*}
%    \exp\((-\sum_{i=1}^N \lambda_a(t_i) \Delta t \cdot \tilde{P}_a\)),
%\end{equation*}
%which is exact in the limit $N \rightarrow \infty$. In this limit the sum turns into an %Riemann
%integral and we %finally
% obtain
%\begin{equation*}
%    \exp\((-\int_{0}^t  \ud \tau \lambda_a(\tau) \cdot \tilde{P}_a \)).
%\end{equation*}
%We can identify the integral of the rate as the average number of encountered receptors
%\begin{equation}
%    m_a(t) = \int_{0}^t \ud \tau \lambda_a(\tau), \quad \lambda_a(t) = \frac{\ud m_a(t)}{\ud t}.
%\end{equation}
%With these identifications, seeing that we assumed the probability of a successful encounter, $\tilde{P}_a$, does not change in time, and by dividing by $\ud t$ the expression for the probability becomes
Taking the $N\to\infty$ limit yields:
\begin{equation}
    H_a(t) = \lambda_a(t) \tilde{P}_a e^{-\int_{0}^{t} dt'\lambda_a(t') \tilde{P}_a},
\end{equation}
where we have used the short-hand notation $\tilde{P}_a = \sum_r P_r f_{r,a}$ for the probability that a randomly chosen receptor recognizes antigen $a$.

%{\color{magenta} The result is somewhat obvious (well except for the thinning and the inhomogeneity, but intuitively it's clear that this should not change the result). Only give reference?}

\section{Convexity of the expected cost}
\label{app:convexity}
In this Appendix we show that the cost function $\av{F}$
is a convex function of its argument $\{P_r\}$ (the receptor distribution).
We start by introducing an alternative expression of $\bar F_a$,
  obtained by integration by parts: %to Eq. \ref{eq:avFp}, that will simplify the following calculations.
%The cost is only defined up to a constant, so we can always choose to
%set $F(0) = 0$.
%Applying the Laplace transform identity $s \mathcal{L}\{f\}(s) =
%\mathcal{L}\{f'\}(s) + f(0)$ \citep{muehlig-2007}, where $f'$ denotes
%a derivative with respect to time $t$, to Eq. \ref{eq:avFp} we obtain
\begin{equation}
\bar F_a  = \int_0^{\infty} \ud m F_a'(m) e^{- m \tilde{P}_a}+F(0).
\end{equation}
We calculate the derivatives of this average cost with respect
to $\tilde P_a$:
\begin{align}
    \frac{\ud \bar{F_a}}{\ud \tilde P_a} &= -\int_0^{\infty} \ud m \,m F_a'(m) e^{- m \tilde{P}_a} \\
    \frac{\ud^2 \bar{F_a}}{\ud \tilde P_a^2} &= \int_0^{\infty} \ud m \, m^2 F_a'(m) e^{- m \tilde{P}_a} \label{eq:ddF}
\end{align}
Since by assumption $F_a'(m)$ is positive,
% non-negative everywhere and positive at least for some values.
the second derivative of $\bar F_a$ with respect to  $\tilde P_a$  is
positive.
This establishes the convexity of $\tilde P_a$ as a
  function of $\tilde P_a$.
%The positivity of the second derivative is a second-order condition establishing convexity of a function, so $\bar F_a$ is convex.
Since $\av{F}=\sum_a Q_a \bar F_a$
(with $Q_a\geq 0$),
% \citep{vandenberghe-2004}
it is a convex function of $\{\tilde P_a\}$. Therefore it is also a convex function of $\{P_r\}$,
as $\{P_r\}$ and $\{\tilde P_a\}$ are linearly related.

\section{Biological motivation of power-law cost functions}
\label{app:costexample}

In the main text we have developed a general framework for discussing the antigen-receptor recognition process.
To fully specify the model we need to choose an effective cost
function $F_a(m)=F_a(t_a(m))$.
In the main text we derive optimal receptor distributions
for a number of effective cost functions, including 
  power-law functions $F(m)=m^\alpha$.
Here we sketch plausible scenarios motivating that choice.
%We motivated the different choices qualitatively and explored the influence of some qualitative features that effective cost functions might exhibit.
%We did not discuss how to explicitly link the effective cost to the choice of a cost function and a time-dependent encounter rate.
%Here 
%We construct a mechanistic model of the antigen population dynamics after infection and show how it can give rise to the effective cost scaling as a power law in the number of encounters.

%{\color{red} Do we want to say this, or do we want the old argument: F~exponential in time, and the number of encounters is exponential in time but with a different exponent. this combination gives a power law?}{\color{magenta} It was my understanding, that what I wrote below was the old argument. It reduces to what you propose, but explains why the exponents might be different.}

Consider an organism being infected with a antigen $a$.
As long as there is no immune reaction, the antigens divide inside its host and thus increase its population size.
If the initial population size is small it is reasonable to assume exponential growth.
%The population size of the antigen at time $t$ is then given as 
%\begin{equation}
%N_a(t)  = N_0 e^{\mu_a t},
%\end{equation}
%where $\mu_a$ is the growth rate.
%In the simplest (well-mixed) case the encounter rate should be proportional to the antigen population
%\begin{equation}
%    \lambda_a(t) = \lambda N_a(t),
%\end{equation}
%where the proportionality constant $\lambda$ is the rate with which a single antigen encounters receptors.

The more antigens there are at the time of the immune reaction the
more damage they can potentially do. Likewise, the more antigens, the
higher the rate of encounters. These two quantities are also expected
to grow exponetially in time:
%Therefore a better proxy for cost than the time of detection might be the antigen population size at detection
\begin{eqnarray} \label{eq:mmcost}
    F_a(t) &=&  F_{a}(0) e^{\nu_a t},\\
\lambda_a(t)&=&\lambda_{a}(0)e^{\nu'_a t}
\end{eqnarray}
%which is exponential in time. As the number of antigens increases exponentially in time, the number of encounters is also exponential in time, albeit with a different rate $\beta$,  
%\begin{equation}
%m\sim e^{\beta t}.
%\end{equation}
The two exponents may be different in general, because the number of pathogenic agents that cause the harm may grow differently than the
number of antigens that can be recognized by the immune system. This
difference could for example come stem from the fact that both the pathogen's
antigenic exposure and its virulence are cooperative effects, and thus
scale as a power of the number of invading individuals.
Using $m_a(t)=\lambda_{a}(0)(e^{\nu'_a t}-1)/\nu'_a$, and
eliminating time $t\approx \ln [m_a/\lambda_a(0)]/\nu'_a$ (for
$t$ large compared to $1/\nu'_a$), we rewrite
the effective cost function in terms of the number of encounters:
\begin{equation}
    F_a(m) =  F_a(0) {\left(\frac{m}{\lambda_a(0)}\right)}^{\frac{\nu_a}{\nu'_a} }\propto m^{\alpha},
\end{equation}
with $\alpha={\nu_a}/{\nu'_a}$.

\section{Analytical optimization}

\subsection{Optimality conditions}
\label{app:optcond}
%Duality theory, an extension of the Lagrangian optimization, can be used to derive a set of necessary conditions for solutions to inequality constrained optimization problems, the so called Karush-Kuhn-Tucker (KKT) conditions \citep{vandenberghe-2004}.
In the following we give optimality conditions for the optimization problem
defined in the main text, which will be used for the following
analytical determination of optimal receptor distributions.
These conditions, called Karush-Kuhn-Tucker conditions
  \citep{vandenberghe-2004},
are derived from a generalization of the
  method of Lagrange multipliers to inequality as well as equality constraints.

The Lagrangian for the optimization problem is
\begin{equation}
    \mathcal{L}(P, \lambda, \nu) = \av{F}(P) + \lambda \((\sum_r P_r - 1\)) - \sum_r \nu_r P_r,
\end{equation}
with
\begin{equation} \label{eq:costApp}
   \av{F}=\sum_a Q_a \bar F_a .
\end{equation}
$\lambda$ is a Lagrange multiplier enforcing the normalization constraint and $\nu_r$ are Lagrange multipliers enforcing the non-negativity constraint.
%Here and in the rest of the appendix we denote vectors by bold symbols.
The optimal $P^*$ is an extremum of this Lagrangian. Thereore the stationarity conditions:
%\begin{equation} \label{eq:stationarity}
%    \nabla \mathcal{L}\((\B P^*, \lambda^*, \B \nu^*\)) = 0
%\end{equation}
\begin{equation}
    \frac{\partial \av{F}}{\partial P_r}\bigg|_{P^*} + \lambda^* - \nu_r^* = 0,
\end{equation}
with
\begin{equation}\label{eq:derAvF}
\frac{\partial \av{F}}{\partial P_r}=\sum_a Q_a\bar F'_a(\tilde P_a) f_{r,a},
\end{equation}
must hold for some value of $\lambda^*$ and $\nu_r^*$ that enforce the constraints.
%Calculating this derivative of the Lagrangian gives us the conditions
%that need to hold for all $r$.
%For the Lagrange multiplier enforcing
The inequality constraint $P_r\geq 0$ further requires that:%we have two further constraints
\begin{align}
    \nu_r^* &\geq 0 \\
    \nu_r^* P_r^* &= 0,
\end{align}
where the second is known as the complementary slackness condition. %\citep{vandenberghe-2004}.
It requires the Lagrange multipliers associated with the
non-negativity to be zero unless the constraint is active, {\em i.e.} unless the corresponding receptor probability is zero.

The three conditions may be reformulated as:
%by recognizing that the $\nu_r^*$ act as slack variables and eliminating them \citep{bertsekas-1999}:
\begin{align}
    \frac{\partial \av{F}}{\partial P_r}\bigg|_{P^*} + \lambda^* & \geq 0 \label{eq:lagrangestationaritynoslack} \\
    \left(\frac{\partial \av{F}}{\partial P_r}\bigg|_{P^*} + \lambda^*\right) P_r &= 0
\end{align}
For all receptors that are present in the optimal repertoire ($P_r^* > 0$) these conditions imply
\begin{equation}
    \frac{\partial \av{F}}{\partial P_r}\bigg|_{P^*} = -\lambda^{*}. \label{eq:optcondinterior}
\end{equation}
If a receptor is not present in the optimal repertoire ($P_r^* = 0$)
then the less stringent condition holds:
\begin{equation}
    \frac{\partial \av{F}}{\partial P_r}\bigg|_{P^*} \geq - \lambda^{*}.
    \label{eq:optcondboundary}
\end{equation}
We note here that ${\partial \av{F}}/{\partial P_r}\leq 0$ 
  (because more receptors always yield a lower cost), so that
  $\lambda^*\geq 0$.

%{\color{magenta} The final result can also be found in \citep{bertsekas-1999}. Only give reference?}

These two conditions can be explained as follows:
if a repertoire is optimal, all changes allowed by the constraints
will lead to a higher cost,
{\em i.e.} moving receptors from one type to another
  will not yield an improvement.
%Replacing a receptor of one type by one of another type should not yield an improvement.
All partial derivatives of the cost with respect to the receptor probabilities should thus be equal to the same value (Eq. \ref{eq:optcondinterior}).
If there are already no receptors of a certain type, {\em i.e.} $P_r = 0$,
we get a less stringent condition.
We can no longer remove receptors away from this type $r$,
but only add some to it, at the expense of other receptor types. %replace other receptors by this receptor.
The increase in cost due to the depletion of these other types
should be higher than the gain of moving them to type $r$.
The partial derivatives of the cost with respect to the receptors 
  that are not
present in the repertoire must thus be larger than the
partial derivatives of the present receptors, which
are given by $-\lambda^*$ (Eq. \ref{eq:optcondboundary}).

\subsection{Solution for uniquely specific receptors}
\label{app:optspecific}

%In this Appendix we show how to {\color{red}optimally
%  bias}{\color{magenta} why colored?} 

%a repertoire of uniquely specific receptors based on the optimality
%conditions introduced above in
We now solve 
 Eqs.~\ref{eq:optcondinterior} and
\ref{eq:optcondboundary} for a repertoire of uniquely specific
receptors (no cross-reactivity).
Eq.~\ref{eq:derAvF} becomes
\begin{equation}
\frac{\partial \av{F}}{\partial P_r} = Q_r \bar F_r'\((P_r\)),
\end{equation}
%where $\bar F_a'=\partial \bar F_a/\partial \tilde P_a$, and 
where we have used the fact that in the absence of cross-reactivity
$\tilde P_a=P_a$.
%First, given some effective cost function the average cost $\bar F$ can be calculated using Eq. \ref{eq:avFp}.
%As the optimality conditions involve the partial derivative of the expected cost with respect to the receptor probabilities, we proceed to calculate these partial derivatives as 
%\begin{equation} \label{eq:avFderiv}
%    \frac{\partial \av{F}}{\partial P_r} = \sum_a Q_a \bar F_a'\((\tilde P_a\)) f_{r,a}.
%\end{equation}
%Here we have used the chain rule and the definition of $\tilde P_a$ (Eq. \ref{eq:coverage}).
%If the receptors are uniquely specific, i.e. $f_{r,a} = \delta_{r,a}$, then this equation simplifies to $\frac{\partial \av{F}}{\partial P_r} = Q_r \bar F_r'\((P_r\))$.
If  all optimal receptor probabilities are positive then we can insert this relationship into Eq. \ref{eq:optcondinterior} to obtain
\begin{equation}
    Q_r \bar F_r'\((P_r^*\)) = -\lambda^*.
\end{equation}
and thus:
%As the last step we solve the above equation for the receptor distribution to obtain
\begin{equation} \label{eq:optspecificpositive}
    P_r^* = h_r\((-\lambda^*/Q_r\)),
\end{equation}
where $h_r = \bar F_r'^{(-1)}$ denotes the inverse function of  $\bar
F'_r$. Since that function $\bar F'_r$  is always negative, $h_r$ must take a
negative argument.

For some cost functions, solving this equation may yield some negative receptor probabilities.
In these cases some of the non-negativity constraints need to be active.
%As we show in the following 
Setting $P_r = 0$ when
Eq.~\ref{eq:optspecificpositive} is negative yields the correct
optimal distribution under the non-negativity constraint. 
We verify that for these $r$,
Eq.~\ref{eq:optcondboundary} { is satisfied} by $P_r=0$, because:
\begin{equation}
    Q_r\bar F_r'\((P_r=0\)) \geq Q_r\bar F_r'[h_r\((-\lambda^*/Q_r\))] = -\lambda^*,
\end{equation}
where we have used the fact that $\bar F_r'$ is a 
  increasing function of its argument (due to the positivity of its
derivative, {\em cf.} Eq. \ref{eq:ddF}), and
$h_r\((-\lambda^*/Q_r\))\leq 0$
% and that thus $\bar F_r'(x) < \bar F_r'(0)$ holds for any $x < 0$.}

In summary, the solution to the optimization problem is 
\begin{equation}
    P_r^* = \max\{{h_r\((-\lambda^*/Q_r\)), 0\}},
\end{equation}
where the value of $\lambda^*$ is fixed by the normalization condition
$\sum_r P_r=1$.

\begin{table}
    \centering
    \begin{tabular}{c|c|c}%\toprule
      $F(m)$ & $\bar F(\tilde P_a)$ & $h(x)$ \\
    \colrule
    $m^\alpha$ & $\Gamma(1+\alpha) / \tilde P_a^\alpha$ &  $\((-x / (\alpha \Gamma(1+\alpha))\))^{\frac{1}{1+\alpha}}$ \\
    $\ln m$ & $\gamma - \ln \tilde P_a$ & $-1/x$ \\
    $1 - \exp \((- \beta m\))$ & $\beta / \((\beta + \tilde P_a\))$ & $\sqrt{-\beta/x} - \beta$ \\
    $\Theta(m-m_0)$ &$\exp\((-m_0 \tilde P_a\))$ & $-\ln(-x/m_0)/m_0$\\
    %\botrule
%        $F(m)$ & $\bar F(P_r)$ & $P_r^*$ \\
%    \colrule
%    $m^\alpha,\, \alpha>0$ & $\Gamma(1+\alpha) / P_r^\alpha$ &  $ c Q_r^{\frac{1}{1+\alpha}}$ \\
%    $\ln m$ & $\gamma - \ln P_r$ & $ c Q_r$ \\
%    $1 - \exp \((- \beta m\))$ & $ \beta / \((\beta + P_r\))$ & $\max\{{c \sqrt{Q_r} - \beta, 0 \}}$ \\
%    $\Theta(m-m_0)$ & $\exp\((-m_0 P_r\))$ & $\max\{{\ln\((Q_r\)) / m_0 - c, 0 \}}$\\
    %     $F(m)$ & $\bar F'(x)$ & $h(x)$ \\
    % %\colrule
    % $m^\alpha,\, \alpha>0$ & $-\alpha \Gamma(1+\alpha) / x^{(1+\alpha)}$ &  $\((-x / (\alpha \Gamma(1+\alpha))\))^{\frac{1}{1+\alpha}}$ \\
    % $\ln m$ & $-1/x$ & $-1/x$ \\
    % $1 - \exp \((- \beta m\))$ & $ - \beta / \((\beta + x\))^2$ & $\sqrt{-\beta/x} - \beta$ \\
    % $\Theta(m-m_0)$ & $-m_0 \exp\((-m_0 x\))$ & $-\ln(-x/m_0)/m_0$\\
    % %\botrule
   \end{tabular}
    \caption{
        Intermediate results in the derivation of the {optimal solution}.
        The first column shows {several choices of} the effective cost function, $F(m)$.
        For these cost functions the second column shows the average
        cost of a pathogenic attack, $\bar F(\tilde P_a)$, and the
        third column shows the inverse of {its derivative}, $h = \((\bar F'\))^{-1}$.
        $\Gamma$ is the Gamma function, $\gamma$ is Euler's constant, $\beta$ and $m_0$ are positive constants.
    }
    \label{tab:biasingderivation}
\end{table}

{In Tab. \ref{tab:biasingderivation} we give the explicit
  expressions of $\bar F_a$ and $h_a$,
  for the particular choices of the cost function $F(m)$ considered in
  the main text.}
%In all these cases
%  $\bar F_a=\bar F(\tilde P_a)$ is a function of $\tilde P_a$ independent of $a$.}
%are special cases of this result.
%For the effective cost functions given in Tab. \ref{tab:costinfluence} the average cost can be derived with little algebra by making use of tables of standard Laplace transforms \citep{muehlig-2007}.
%Intermediate results for $\bar F'$ and $h$ are given in Tab.~\ref{tab:biasingderivation}.
%For ease of notation we then give the results in terms of a constant $c$, that is a multiple on $\lambda$, getting rid of unimportant multiplicative factors.

\subsection{Solution for cross-reactive receptors}
\label{app:optcross}
%In this section we give details of the analytical calculations we performed to obtain the results in the presence of cross-reactivity that are presented in the {\it Cross-reactivity dramatically limits optimal repertoire diversity} section of the main text.
{The previous results can be generalized to cross-reactive
  receptors in a continuous space, using Fourier transforms. This
  generalization will lead up to the results presented in the {\it
    Cross-reactivity dramatically limits optimal repertoire diversity}
  section of the main text, and notably the Gaussian case discussed therein.}

\subsubsection{Deconvoluting the optimality conditions in Fourier space}
We consider a continuous receptor-antigen space %for ease of notation
and we assume a translation invariant cross-reactivity function $f_{r,a} = f(r-a)$.
We write the optimality condition Eq. \ref{eq:optcondinterior}
\begin{equation}\label{eq:optcontinoptexp}
    \int \ud p \; Q(a) \bar F'\left[\tilde P^*(a)\right] f(r-a) = -\lambda^*,
\end{equation}
where in continous space the coverage is defined as:
\begin{equation}
    \tilde P(a) = \int \ud r \; P(r) f(r-a).
\end{equation}
We notice that both expressions involve integrals, which are convolutions with the cross-reactivity kernel.
Since the convolution of a constant is also a constant, a solution of 
\begin{equation}
    Q(a) \bar F'\((\tilde P^*(a)\)) = -\lambda',\quad\textrm{with }\lambda'>0,
\end{equation}
is also a solution of Eq.~\ref{eq:optcontinoptexp}.
As in the case of uniquely specific receptors, we can solve this equation for $\tilde P^*(a)$:% to obtain
\begin{equation} \label{eq:Ptildestar}
    \tilde P^*(a) = h\[[-\lambda'/Q(a)\]],
\end{equation}
where $h = \bar F'^{(-1)}$ as in \ref{eq:optspecificpositive}.
{If there was no} cross-reactivity, there {would be} no difference between $P$ and
$\tilde P$, {and we would be done.}
Here we need to perform a deconvolution to obtain the optimal receptor distribution $P$ from the optimal coverage $\tilde P$.
We do so in Fourier space, where the convolution turns into a product. %\citep{muehlig-2007}
Deconvolution is therefore much simpler in Fourier space as it corresponds to a division
\begin{equation} \label{eq:fourierconvolution}
    \mathcal{F}[\tilde P] = \mathcal{F}[P] \mathcal{F}[f] \quad \Leftrightarrow \quad \mathcal{F}[P] = \mathcal{F}[\tilde P] / \mathcal{F}[f],
\end{equation}
where we have defined the Fourier transform of a function $g(x)$ as
$\mathcal{F}[g](k) =\int_{-\infty}^{\infty} \ud x g(x) e^{ikx}$.
To calculate the optimal receptor distribution we insert
Eq. \ref{eq:Ptildestar} into Eq. \ref{eq:fourierconvolution} and
perform an inverse Fourier transform $\mathcal{F}^{-1}[\tilde g](x) =
({1}/{2\pi})\int_{-\infty}^{\infty} \ud k \tilde g(k) e^{-ikx}$ to obtain
\begin{equation} \label{eq:optfourier}
    P^* = \mathcal{F}^{-1}\left[\mathcal{F}[h\((-\lambda'/Q\))] / \mathcal{F}[f]\right].
\end{equation}
{This result is only valid as long as the above quantity is
  positive and normalizable, as we shall see below.}

\subsubsection{The Gaussian case}
\label{app:gaussian}
In this section we apply the general results of the previous section to a concrete example.
In order to find the optimal receptor distribution analytically we use Eq. \ref{eq:optfourier}, we assume the antigen distribution and cross-reactivity function are Gaussian
\begin{align}
    Q(a) &= \frac{1} {\sqrt{2 \pi \sigma_Q^2}} \exp\left(-a^2/{2\sigma_Q^2}\right), \\
    f(r-a) &= \exp\[[-(r-a)^2 / 2 \sigma^2\]],
\end{align}
 and %for concreteness
we take
 \begin{align}
    F(m) &= m^\alpha.
\end{align}
Inserting $h$ from Tab.~\ref{tab:biasingderivation} into
Eq.~\ref{eq:optfourier} 
%and using the linearity of the Fourier transform
allows us to write
\begin{equation}
P^* \propto \mathcal{F}^{-1}\[[\mathcal{F}[Q^\frac{1}{1+\alpha}] / \mathcal{F}[f]\]]
\end{equation}
as an equivalent equation determining the optimal repertoire.
We can calculate the modified antigen distribution as
\begin{equation}
    Q(a)^\frac{1}{1+\alpha} \propto \exp\((-\frac{a^2}{2(1+\alpha)\sigma_Q^2}\)).
\end{equation}
The Fourier transform of a Gaussian function of variance $\sigma^2$ is a Gaussian function of variance $1/\sigma^2$ \citep{muehlig-2007}. Therefore we have
\begin{align}
    \mathcal{F}[Q^\frac{1}{1+\alpha}](q) &\propto \exp\[[-{(1+\alpha)\sigma_Q^2q^2}/{2}\]], \\
    \mathcal{F}[f](q) &\propto \exp\[[-{\sigma^2a^2}/{2}\]],
\end{align}
from which
\begin{equation}
    \mathcal{F}[Q^\frac{1}{1+\alpha}] / \mathcal{F}[f] \propto \exp\left\{-{[(1+\alpha)\sigma_Q^2-\sigma^2]q^2}/{2}\right\}
\end{equation}
follows.
Taking the inverse Fourier transform {and normalizing}, we obtain
\begin{equation}
    P^*(r)= \frac{1}{\sqrt{2 \pi [(1+\alpha)\sigma_Q^2 - \sigma^2]}} \exp\((-\frac{r^2}{2[(1+\alpha)\sigma_Q^2-\sigma^2]}\)).
\end{equation}
%We normalize this result to obtain the optimal receptor distribution
%\begin{equation}
%    P^*(r) = \frac{\exp\((-\frac{r^2}{2 ((1+\alpha)\sigma_Q^2 - \sigma^2)}\))}.
%\end{equation}

Normalization is only possible for $\sigma < \sigma_Q\sqrt{1+\alpha} \equiv \sigma_c$.
In the limit $\sigma \rightarrow \sigma_c$ the Gaussian converges to a Dirac delta function.
Intuition suggests that a Dirac delta function centered on the peak position should remain optimal for further increases in $\sigma$.
To prove this assertion we note that a Dirac delta function is zero everywhere, except in one point.
Since all but one receptor probabilities are at the boundary defined by the non-negativity constraints, we only need to check Eq.~\ref{eq:optcondboundary}.
We compute the {left-hand side} of Eq.~\ref{eq:optcontinoptexp} as a function of $r$
\begin{equation}
    \begin{split}
        &\int \ud p \; Q(a) \bar F'[\tilde P^*(a)] f(r-a) \\
        &\quad \propto - \exp\left\{\frac{- r^2 [\sigma^2 - (1+\alpha) \sigma_Q^2]}{2\sigma^2(\sigma^2-\alpha\sigma_Q^2)}\right\},
    \end{split}
\end{equation}
and note that it has a minimum for $r=0$.
This shows that the partial derivatives of the expected cost at $r \neq 0$ are greater than at $r = 0$, implying that Eq. \ref{eq:optcondboundary} holds.

The cost of the optimal repertoires as a function of the cross-reactivity width $\sigma$ is given by
\begin{equation}
    \av{F}(P^*) = %(2 \pi)^{\frac{1+\alpha}{2}}\sigma\times
    {\left(\frac{\sigma_Q}{\sigma}\right)}^\alpha
   \begin{cases}
        (1 + \alpha)^{\frac{1+\alpha}{2}} & \text{if } \sigma < \sigma_c,\\
       \frac{(\sigma/\sigma_Q)^{\alpha}}{\sqrt{1 - \alpha (\sigma_Q/\sigma)^2}} & \text{otherwise}.
    \end{cases}
\end{equation}
Both expressions give the same cost at the transition $\sigma =
\sigma_c$. After multiplying by $(\sigma/\sigma_Q)^\alpha$ to compare
at constant recognition capability $\int f=\sqrt{2\pi}\sigma$, this
expression is constant for $\sigma<\sigma_c$, and grows for $\sigma>\sigma_c$.

\subsubsection{{General} argument for peakedness}
\label{app:mathpeaks}

%\todo{AM: this section needs to be updated}
A simple argument can {help understand} why cross-reactivity generically leads to peaked optimal solutions.
The convolution with a kernel is a smoothening operation, {represented by a low-pass filter in the
Fourier domain}.
The optimal solution in the absence of the non-negativity constraints
requires that $\tilde P_a = h(Q_a)$.
As $\tilde P_a$ is the low-passed filtered version of $P_r$, the
high-frequency components of $h(Q_a)$ will be magnified {by the
  deconvolution}.
{These high-frequency wiggles can lead to negative values of $\mathcal{F}^{-1}[h(Q_a)]$,
  which are not allowed, leading to set many values of $P(r)$ to zero.}
%The non-negativity does not allow for large amplitude wiggles in
%$P_r$, which constrains the optimal solution. % many of the non-negativity constraints will thus be active.
This effects results in a peaked solution.
Because the size of the cross-reactivity kernel is inversely
proportional to the cutoff frequency in the Fourier domain, we expect
the spacing of the peaks to be related to the size of the cross-reactivity kernel.

\section{Numerical optimization}
\label{app:numerics}
We numerically minimize the cost function subject to the normalization
and non-negativity constraints by using a fast projected gradient
algorithm. In the following we provide details on this numerical
algorithm. To facilitate notations let us define the 
function {to minimize as} $g(x)$, {where $x$ is a vector in a Euclidean space}, and the convex set $C$ defined by the constraints. In these notations the problem we want to solve can be stated as
\begin{equation}
    \min_{x \in C} g(x).
\end{equation}

Given an arbitrary starting point $x_0 \in C$ the algorithm performs the following iterative procedure:
\begin{align}
    y^{k+1} &= x^k + \omega^k \((x^k - x^{k-1}\)), \\
    x^{k+1} &= \mathcal{P}\((y^{k+1} - s^k \nabla g\((y^{k+1}\))\)),
\end{align}
where $\nabla$ denotes
the gradient.
Here $\mathcal{P}$ denotes a projection onto $C$, $\omega^k$ is an extrapolation step size and $s^k$ is the step size taken in the direction of the gradient.
The extrapolation step size has to be chosen carefully to ensure the faster convergence of this method with respect to an ordinary gradient method.
Following \cite{boyd-2013} we use
\begin{equation}
\omega_k = \frac{k}{k+3}.
\end{equation}
The step size $s$ is determined by backtracking \cite{teboulle-2009}:
we iteratively decrease $s$ {by multiplication by $\beta<1$} until $g(z) \leq g(y^k) +
(x-y)\cdot\nabla g(y^k) + \frac{1}{2s}( z-y )^2$, where $x \cdot y$ denotes the inner dot product between $x$ and $y$,
and $z = \mathcal{P}(y^k - s \nabla g(y^k))$.
In practice we determine $s$ in this way at the first step of the
optimization and then keep it fixed based on this initial estimate.

%The step size is determined by backtracking \cite{teboulle-2009} for
%some $\beta < 1$:
%{\TM [TM- I don't understand these three lines; also parenthesis
%  missing in first eq.]}
%\begin{align}
%    &\text{Let} \; z = \mathcal{P}(y^k - s \nabla g(y^k) \\
%    &\text{break if} \; g(z) \leq g(y^k) +  (x-y)\cdot\nabla g(y^k) + \frac{1}{2s}( z-y )^2 \\
%    &\text{Update} \; s = s \beta
%\end{align}
%{\TM where $\nabla$ denotes the gradient, and $x\cdot y$ is the inner
%  dot product between $x$ and $y$.}
%In practice we saw that it suffices to determine $s$ in this way in the beginning and to then keep a fixed step size.

The projection of a point onto a convex set is defined by the following quadratic programming problem:
\begin{equation}
    \mathcal{P}(y) = \argmin_{x \in C} \frac{1}{2} ( x-y)^2.
\end{equation}
If the convex set is a simplex as is the case for our problem, there fortunately exist efficient algorithms for solving this problem.
We use the algorithm described in \cite{chandra-2008}.

To stop the iteration one needs to define a suitable stopping criteria.
As the problem is convex we can establish a lower bound for the cost by solving a linear programming problem as follows:
\begin{equation}
    g_{lb} = g(x^k) + \min_{x \in C}  \left[(x - x^k)\cdot \nabla g(x^k) \right] \leq g(x^*).
\end{equation}
The linear programming problem $\bar{x}^k = \argmin_{x \in C} \nabla g(x^k)^T (x - x^k)$ is solved explicitly \citep{bertsekas-1999} by
\begin{equation}
    \bar{x}^k = e_{i^*}, \; i^* = \argmin_i (\nabla g(x^k))_i,
\end{equation}
where $e_{i}$ denotes the ith unit vector.
We can use this lower bound to define a stopping criterion for the numerical optimization
\begin{equation}
    \frac{g(x^k) - g_{lb}}{g_{lb}} < \epsilon.
\end{equation}
For all reported numerical results we have chosen $\epsilon = 10^{-8}$.

The discretization steps used in the figures are listed below:

\begin{center}
\begin{tabular}{c|c}
Step & Figure \\
\hline\hline
$0.25 \sigma$ & 5, \ifthenelse{\equal{\ispnas}{true}}{S5}{\ref{fig:sooapp}} \\
$0.1 \sigma$ & 3, \ifthenelse{\equal{\ispnas}{true}}{S2}{\ref{fig:kappascaling}}, \ifthenelse{\equal{\ispnas}{true}}{S3}{\ref{fig:kernels}}, \ifthenelse{\equal{\ispnas}{true}}{S4}{\ref{fig:correlation}} \\
$0.05 \sigma$ & 4
\end{tabular}
\end{center}

\section{Stable fixed point of population dynamics minimizes corresponding cost function}
\label{app:popdynfixedpoint}
In this section we show that the {stable} fixed point
$\{N_r^*\}$ of the population dynamics:
\begin{equation} \label{eq:popdynApp}
\frac{\ud N_r}{\ud t} = N_r \[[\sum_a Q_a A\((\sum_r N_r f_{r,a}\)) f_{r,a} - d \]]
\end{equation}
gives a probability distribution $P_r = N_r / N_{\rm tot}$ (with
$N_{\rm tot}=\sum_r N_r$) that
minimizes the cost $\av{F}$. {For this correspondence to be
  exact,} the availability function of the dynamics and the effective
cost function of the optimization {must be related by:
\begin{equation} \label{eq:condlyapunovApp}
    A(\tilde N_a)=-c'\bar F'(\tilde N_a/N_{\rm st}),
\end{equation}
where $\tilde N_a=\sum N_r f_{r,a}$, and 
$N_{\rm st}$ is the total number of receptors $N_{\rm tot}$ at
the fixed point.}

{A fixed point is characterized by ${\ud N_r}/{\ud t}= 0$.}
If $N_r > 0$, this translates into
\begin{equation}
\sum_a Q_a A\((\sum_r N_r f_{r,a}\)) f_{r,a} - d = 0.
\end{equation}
Using the correspondence between availability and cost function given by Eq.~\ref{eq:condlyapunovApp} we rewrite this condition as
\begin{equation}
    \sum_a Q_a \bar F'\((\tilde P_a\)) f_{r,a}  = - c' d,
\end{equation}
which is equivalent to the optimality condition Eq.~\ref{eq:optcondinterior}, with the identification $\lambda^* = c' d$.

For $N_r = 0$ we need to work a bit harder to show that the optimality condition at the boundary Eq.~\ref{eq:optcondboundary} is satisfied.
Here the key assumption establishing the minimization of the cost function is the stability of the fixed point.
%the dynamical model.
{A fixed point is stable if the real parts of the Jacobian's eigenvalues are all negative.}
%For a fixed point to be stable, all local perturbations need to decay and linear stability analysis thus requires the real parts of all eigenvalues of the Jacobian at the fixed point to be negative. %\cite{strogatz-2006}.
{The Jacobian reads:}
\begin{equation}
    \begin{split}
         J_{r,r'} = &\delta_{r,r'} \(( \sum_a Q_a A\((\sum_{r'} N_{r'} f_{r', a}\)) f_{r,a} - d\)) \\
        &+ N_r \sum_a Q_a A'\((\sum_{r'} N_{r'} f_{r', a}\)) f_{r,a} f_{r',a}.
    \end{split}
\end{equation}
We remark that for $N_r = 0$ the $r^{\textrm{th}}$ row of the Jacobian is non-zero only on the diagonal.
That value on the diagonal is an eigenvalue of the Jacobian and
must be negative:
\begin{equation}
\sum_a Q_a A\((\sum_{r'} N_{r'} f_{r',a}\)) f_{r,a} - d < 0,
\end{equation}
Again we replace $A\((\sum_r N_r f_{r,a}\))$ by $- \bar F_a'\((\tilde P_a\))$ according to Eq.~\ref{eq:condlyapunovApp} to obtain
\begin{equation}
\sum_a Q_a \bar F'\((\tilde P_a\)) f_{r,a}  > - c' d,
\end{equation}
which is equivalent to the optimality condition at the boundary
Eq.~\ref{eq:optcondboundary}, provided that $\lambda^* = c' d$.

\section{Cost function as a Lyapunov function of the dynamics}
\label{app:popdynlyapunov}
{Here we show rigorously that, when the availability function is scale
  invariant, as in the case for the simple cost function
  $F(m)=m^\alpha$, the dynamics must converge towards a fixed
  point. This fixed point is unique and corresponds to the optimal of the cost $\av{F}$, as we
  have shown in the previous section.}

{$A(x)$ is scale invariant if there exists a function $v$ such
  that} $A(\gamma x)= v(\gamma) A(x)$. In this case {we will see that} the changes of relative
frequencies $P_r$ in the repertoire over time {only} depend on the
total number of receptors {through a prefactor}.
%In this case the temporal evolution of the total population size does not have an influence on the dynamics of relative changes between different receptor types.
Below we derive the equations governing this dynamics and will then prove that this dynamics is assured to converge to a stable fixed point.
We do so by showing that the dynamics {admits the expected cost
  $\av{F}$} as a Lyapunov function, {\em i.e.} a function that continually decreases under the dynamics. %\citep{strogatz-2006}

For ease of notation we rewrite Eq.~\ref{eq:popdynApp} as: 
%we introduce the following dynamical equation of which the population dynamics we consider is a subtype:
\begin{equation}
    \frac{\ud N_r}{\ud t} = N_r [\pi_r(N) -d],
\end{equation}
where $N$ is a short-hand for $\{N_r\}$, and
$\pi_r=\sum_a Q_a A\((\sum_r N_r f_{r,a}\)) f_{r,a}$ is the growth rate of receptor type $r$.
% is the derivative $\pi_r = \partial \pi / \partial N_r$ of a functional $\pi(\{N_{r'}\})$.
%How do the relative frequencies of the competing species change over
%time? We have $P_r = N_r/N$, where $N = \sum_r N_r$ is the total population size.
%We calculate the temporal derivative of the relative frequencies to be
{The relative frequencies $P_r=N_r/N_{\rm tot}$ evolve
  according to:}
\begin{align}\label{eq:dyndreqchange}
    \frac{\ud P_r}{\ud t} &= \frac{1}{N_{\rm tot}} \frac{\ud N_r}{\ud t} -
    \frac{N_r}{N_{\rm tot}^2}\frac{\ud N_{\rm tot}}{\ud t} \\
             &= P_r \[[\pi_r(N) - \sum_{r'} P_{r'} \pi_{r'}(N) \]].
\end{align}
If $A$ is scale invariant, so is $\pi_r$ and $\pi_r(N)=\pi_r(N_{\rm tot}P)
= v(N_{\rm tot}) \pi_r(P)$. 
%The dynamical equations for the change in the
%relative frequencies of the competing species
%(Eq.~\ref{eq:dyndreqchange})
Then the equations further simplify to
\begin{align}
     \frac{\ud P_r}{\ud t} &= v(N_{\rm tot}) P_r \[[\pi_r(P) - \sum_{r'} P_{r'} \pi_{r'}(P) \]], \\
             &= v(N_{\rm tot}) P_r \((\pi_r - \bar \pi\)), \label{eq:replicator}
\end{align}
where $\bar \pi = \sum_r P_r \pi_r$.

{We can now write how the expected cost $\av{F}$ evolves in time:
\begin{align}
    \frac{\ud \av{F}}{\ud t} &= \sum_r \frac{\partial \av{F}}{\partial P_r} \frac{\ud P_r}{\ud t} \\
&=v(N_{\rm tot})\sum_r P_r \left[\sum_a Q_a\bar F'_a(\tilde P_a) f_{r,a}\right] \((\pi_r - \bar \pi\))\\
& =-\frac{v(N_{\rm tot})}{c'} \sum_r P_r \left[\sum_a Q_a A(N_{\rm st}\tilde P_a) f_{r,a}\right] \((\pi_r - \bar \pi\))\\
& =-\frac{v(N_{\rm tot})v(N_{\rm st})}{c'} \sum_r P_r \pi_r \((\pi_r - \bar \pi\))\\
                       &= -\frac{v(N_{\rm tot})v(N_{\rm st})}{c'} \sum_r P_r \((\pi_r - \bar{\pi}\))^2 \leq 0.
\end{align}
%Eq.~\ref{eq:condlyapunovApp}.
This proves that the cost always decreases with time, {\em i.e.}
  is a Lyapunov function of the dynamics. Therefore the
  dynamics will reach a stable fixed point at steady state, which is garanteed to be
  the global minimum of the expected cost $\av{F}$.
}
% We can now show that the functional $\Pi$ is a Lyapunov function of the dynamics

% \begin{align}
%     \frac{\ud \pi}{\ud t} &= \sum_r \frac{\partial \pi}{\partial P_r} \frac{\ud P_r}{\ud t} \\
%                         &= c \(( \sum_r P_r \pi_r^2 - \bar{\pi}^2 \)) \\
%                         &= c \sum_r P_r \((\pi_r - \bar{\pi}\))^2 \leq 0
% \end{align}

%Neglecting the varying time constant $c$ Eq. \ref{eq:replicator} is known as a replicator equation \citep{sigmund-1998}.
%Replicator equations with linear $f_r$ have been widely studied \citep{opper-1989, tokita-2004} and 
%the existence of a Lyapunov function is an established result for equations of this type.

%{\color{magenta} We could again not give the Lyapunov proof, but instead refer to the literature.}

%\clearpage

\bibliographystyle{pnas}
\bibliography{optimmune,thierry,byhand,vijay}

\end{document}